\documentclass[twocolumn,aps,superscriptaddress,prd,10pt,showpacs,preprintnumbers,nofootinbib,eqsecnum,nolongbibliography]{revtex4-2}
\usepackage{graphicx,slashed,bbold,subfigure,amsmath,amssymb,enumerate,color,xcolor,dcolumn,amsthm}
\usepackage{multirow}
\usepackage{soul}
\usepackage{natbib} 
\usepackage[normalem]{ulem}
\usepackage[colorlinks=true,citecolor=blue,linkcolor=blue,urlcolor=blue]{hyperref}
\usepackage{array}
\setcitestyle{square, numbers, comma, sort&compress}

\newcommand\Rsout{\bgroup\markoverwith{\textcolor{red}{\rule[0.5ex]{2pt}{0.4pt}}}\ULon}

\def\lsim{\stackrel{\scriptstyle <}{\phantom{}_{\sim}}}
\def\gsim{\stackrel{\scriptstyle >}{\phantom{}_{\sim}}}

\def\mev{{\, \rm MeV}}
\def\fm{{\, \rm fm}}
\def\km{{\, \rm km}}

\allowdisplaybreaks

\begin{document}
\title{Relativistic mean-field model with density- and isospin-density-dependent couplings}

\author{Gabriel Frohaug}
\affiliation{Department of Physics, University of Houston, Houston, TX 77204, USA}

\author{Konstantin Maslov}
\email{kmaslov@central.uh.edu}
\affiliation{Department of Physics, University of Houston, Houston, TX 77204, USA}

\author{Veronica Dexheimer}
\affiliation{Center for Nuclear Research, Department of Physics, Kent State University, Kent, OH 44242 USA}

\author{Joaquin Grefa}
\affiliation{Department of Physics, University of Houston, Houston, TX 77204, USA}
\affiliation{Center for Nuclear Research, Department of Physics, Kent State University, Kent, OH 44242 USA}

\author{Johannes Jahan}
\affiliation{Department of Physics, University of Houston, Houston, TX 77204, USA}

\author{Claudia Ratti}
\affiliation{Department of Physics, University of Houston, Houston, TX 77204, USA}

\author{Tulio E. Restrepo}
\affiliation{Department of Physics, University of Houston, Houston, TX 77204, USA}

\begin{abstract}
	We present a new hadronic EoS with hyperons built within the relativistic mean-field (RMF) formalism with baryon-density- and isospin-density-dependent couplings. Motivated by microscopic calculations showing density- and isospin-asymmetry-dependence of self-energies, we implement a new form for the baryon-meson couplings. The parameters for the couplings are constrained by a Bayesian analysis, which anchors the model to nuclear saturation properties, chiral effective field theory ($\chi$EFT) predictions for pure neutron matter, heavy-ion collision data, and HALQCD-based hyperon potential calculations at 3-momentum $|\mathbf{k}|=0$ in both isospin-symmetric and pure neutron matter. The resulting EoS satisfies neutron star mass-radius constraints from NICER and GW170817, providing another way to address the hyperon puzzle. The low-density part of the EoS is described via nuclear statistical equilibrium with modern mass tables (AME20/FRDM12, 8244 nuclei), providing a novel and complete general-purpose EoS for astrophysical simulations.
\end{abstract}

\maketitle

\section{Introduction}
The equation of state (EoS) of hot and dense strongly interacting matter is a cornerstone problem of modern nuclear physics~\cite{MUSES:2023hyz}. It is required for understanding the structure of neutron stars (NS), numerical simulations of core-collapse supernovae and binary neutron-star mergers, as well as for the hydrodynamic modeling of heavy-ion collisions (HICs). The multimessenger observation of gravitational waves and electromagnetic signals from the double neutron-star merger GW170817~\cite{LIGOScientific:2017vwq,LIGOScientific:2018cki,LIGOScientific:2017ync} has opened a new era in astronomy, allowing to probe matter in conditions unreachable on Earth or in isolated neutron stars. Recent unprecedented results on neutron star radii from NASA's NICER mission~\cite{Miller:2019cac, Riley:2019yda,Riley:2021pdl, Miller:2021qha} provide additional requirements for the NS EoS.
All these new experimental advances are extensively used to constrain the behavior of hot and dense strongly interacting matter. Complementary to that, properties of nearly isospin-symmetric nuclear matter (ISM) are being investigated based on HICs at RHIC, LHC, and future facilities such as FAIR and J-PARC \cite{Sorensen:2023zkk,Miake:2021rgy}; besides, there are plans to measure isospin-asymmetric matter at FRIB and RIKEN \cite{Sorensen:2023zkk}.

On the theoretical side, the three main first-principle approaches to strong interactions are perturbative QCD (pQCD)~\cite{Haque:2014rua}, lattice QCD (LQCD)~\cite{Borsanyi:2013bia, HotQCD:2014kol, Bazavov:2017dus,Ratti:2018ksb} and chiral effective field theory ($\chi$EFT)~\cite{Hebeler:2009iv,Tews:2012fj,Drischler:2021kxf}. pQCD is only valid in the weakly interacting regime, which for QCD only takes place at extremely high energies. Due to the sign problem in LQCD \cite{Nagata:2021ugx}, its applicability is limited in baryon chemical potential ($\mu_B$) or baryon density ($n_B$), both related to the excess of baryons over antibaryons. Even using the most advanced expansion schemes~\cite{Borsanyi:2021sxv,Abuali:2025tbd}, reliable LQCD results are reserved to the ratio $\mu_B/T \lsim 3.5$. The validity of $\chi$EFT is also limited to low temperatures and density in the range $n_B \sim (1-2)\,n_0$, where $n_0 \simeq 0.16\fm^{-3}$ is the nuclear saturation density~\cite{Gross-Boelting:1998qhi,PREX:2021umo}.  However, the simulation of astrophysical transients requires knowledge of the EoS in a wide range of temperatures $T = (0 \textendash 100)\mev$, densities $n_B = (10^{-11} \textendash 10) \, n_0$, and electron fraction $Y_e = n_e/n_B = (0 \textendash 0.6)$~\cite{Typel:2013rza,Oertel:2016bki}, well outside the range of applicability of pQCD (at least directly~\cite{Komoltsev:2021jzg,Mroczek:2023zxo}), or controlled lattice QCD (LQCD) approximations \cite{Bollweg:2022fqq,Borsanyi:2021sxv}.

Phenomenological effective models provide a practical framework to incorporate experimental constraints and QCD symmetries across the full thermodynamic range relevant to astrophysical applications. Such models include $\sigma-\omega$ or Walecka-type relativistic mean-field (RMF) models, where $\sigma$ can be associated with the $f_0(500)$ meson~\cite{Walecka1986,Ring:1996qi,Dutra:2014qga,Lopes:2024bvz,Abhishek:2023uha}, Skyrme and Gogny models~\cite{Chabanat:1997qh, Stone:2006fn, Dutra:2012mb,Sun:2023xkg}, or chiral models~\cite{Dexheimer:2008ax,Dexheimer:2008cv,Fraga:2023wtd} for hadronic matter, and (Polyakov) -- Nambu -- Jona-Lasinio-based~\cite{Hatsuda:1994pi, Ratti:2005jh, Blaschke:2005uj, Dexheimer:2009hi,Motornenko:2019arp,Celi:2025zmn} and similar relativistic density functional approaches~\cite{Kaltenborn:2017hus, Ivanytskyi:2022oxv} for describing the possibility of quark matter cores of massive NSs. Recently, substantial progress was made to build model-agnostic approaches, such as speed-of-sound parameterizations~\cite{Chatziioannou:2018vzf,Tews:2018kmu, OBoyle:2020qvf,Tan:2020ics,Koehn:2024set}, piecewise polytropes~\cite{Read:2008iy,Steiner:2015aea,Raithel:2016bux} and the meta-modeling approach~\cite{Margueron:2017eqc, Margueron:2017lup, Char:2025zdy}. However, from these model-independent approaches, only meta-modeling allows to investigate not only the bulk EoS, but also the matter composition, which is important for understanding detailed properties of neutron star matter, such as transport parameters~\cite{Most:2022yhe}, neutron star cooling curves~\cite{Yakovlev:2000jp,Blaschke:2004vq,Page:2004fy,Sagun:2023rzp}, and intermediate-density phase transitions~\cite{Mondal:2023gbf,Ferreira:2021yje}.

From the point of view of many-body theory, RMF models can be interpreted as the Hartree approximation to the meson-exchange theory, supplemented by a different behavior of scalar and vector potentials under Lorentz transformations that eventually leads to nuclear saturation and large spin-orbit coupling~\cite{Serot:1984ey}. 
The simplest RMF models with constant couplings cannot describe the in-medium modifications to the interaction properties~\cite{Walecka1986}. Additionally, more flexibility is required to simultaneously describe the vast number of available experimental constraints. The coupling constants of such models are typically fitted to reproduce the bulk properties of nuclear matter at saturation, and therefore phenomenologically include the effects of more complicated many-body diagrams including exchange terms, $N$-body correlations and dynamical screening of the interaction.
A common strategy to resolve this is to introduce phenomenological modifications of hadron couplings and/or masses either through their dependence on the baryon density~\cite{Typel:1999yq, Typel:2009sy} or the scalar fields~\cite{Kolomeitsev:2004ff,Gomes:2014aka}. Alternative ways of introducing more flexibility into the models include non-linear couplings~\cite{Horowitz:2000xj,Dexheimer:2020rlp,Pradhan:2022txg}, tensor couplings~\cite{Alberto:2004kb,Typel:2024myq}, or non-linear derivative terms in the Lagrangian~\cite{Chorozidou:2024gyy,Kochankovski:2025thf}.

Available Dirac--Br\"uckner--Hartree-Fock (DBHF) calculations~\cite{Brockmann:1990cn,Katayama:2013zya} predict that all the scalar, vector, vector-isovector, and scalar-isovector self-energies depend noticeably on both the total baryon density and the isospin asymmetry of the medium. The former realization has motivated the development of RMF models with density-dependent (DD) couplings. The first appearance of such kind of dependence (although on scalar fields instead of explicitly on density) in relativistic models was in~\cite{Delfino:1995ea}, where a generalization of the Walecka \cite{Walecka1986} and Zimanyi and Moszkowski~\cite{Zimanyi:1990np} models appeared in the form of derivative baryon-meson couplings. Introduced in~\cite{Typel:1999yq}, the DD coupling formalism was inspired by microscopic DBHF results and allowed to provide an adequate description of nuclear matter properties and finite nuclei. The DD2 parameterization~\cite{Typel:2009sy} became widely used due to its successful description of nuclear matter and NS properties. However, modern data on compact star radii and tidal deformabilities have excluded the original DD2 parameterization. Recent studies have proposed alternative DD parameterizations with improved description of NS properties~\cite{Malik:2022zol,Huang:2024yjx,Boukari:2024wrg}.

In addition to the different possible couplings, the EoS of nuclear matter is strongly dependent on the appearance of multiple new degrees of freedom (besides nucleons and electrons). The composition of the neutron-star inner core is currently unknown, with multiple possibilities discussed in the literature. The appearance of hyperons~\cite{Glendenning:1984jr} or $\Delta$-isobars~\cite{Schurhoff:2010ph,Drago:2014oja} is widely studied within hadronic models, as well as the possibility of pion~\cite{Migdal:1973zm,Sawyer:1973fv,Migdal:1990vm}, kaon~\cite{Kaplan:1986yq,Brown:1993yv}, or $\rho$-meson~\cite{Voskresensky:1997ub,Kolomeitsev:2004ff,Kolomeitsev:2017gli} condensation in the cores of massive neutron stars. Another complication arises due to the appearance of nuclear clusters at low densities, which significantly affect neutrino transport properties in simulations of supernovae and BNS mergers~\cite{Hempel:2009mc,Furusawa:2022ktu}. The main ways of treating them are either the single-nucleus approximation (SNA)~\cite{Lattimer:1991nc,Shen:1998gq} or the nuclear statistical equilibrium (NSE) approach~\cite{Hempel:2009mc, Furusawa:2017auz} with in-medium modifications of nuclear properties; we adopt the latter approach in this work.

The appearance of new degrees of freedom softens the NS EoS, which leads to a decrease of the predicted maximum NS mass~\cite{Glendenning:1997wn,Weissenborn:2011kb,Maslov:2015wba}. If one calibrates the model to describe the hyperon potentials at nuclear saturation density $U_\Lambda(n_0) \simeq -30\mev$, $U_\Sigma(n_0) \simeq (+10\textendash50)\mev$, $U_\Xi(n_0) \simeq (-5\textendash20)\mev$~\cite{gal2016,Tolos_2020}, 
then the hyperon appearance is inevitable at densities $n\simeq(2-3)\,n_0$~\cite{Glendenning:1991es}. The resulting decrease of the predicted neutron star mass is so strong that it has ruled out the majority of models available in the 2010s, a problem dubbed as the ``hyperon puzzle''~\cite{Bombaci:2016xzl,Vidana:2018bdi}. Recent works within phenomenological models show that it is possible to remedy the hyperon softening effect on the NS EoS by introducing additional repulsion in the hyperon-nucleon and hyperon-hyperon interactions~\cite{Schaffner-Bielich:2000igu,Oertel:2012qd,Banik:2014qja}.
Apart from the modification of the hyperon in-medium interactions, other ways of solving this issue include a transition to quark matter~\cite{Annala:2019puf} or quarkyonic matter~\cite{McLerran:2018hbz, Zhao:2020dvu, Margueron:2021dtx, Moss:2024uam}, or considering momentum-dependent potentials~\cite{Chorozidou:2024gyy, Kochankovski:2025thf}.

Hypernuclear experiments can provide information on hyperon interactions with nearly isospin-symmetric matter only. For the typical isospin asymmetry of a neutron star, however, one has to rely on microscopic calculations to constrain the phenomenological model predictions. 
The input baryon-baryon interactions have been recently obtained from the Hadrons to Atomic nuclei from Lattice QCD (HALQCD) Collaboration~\cite{Nemura:2011uzb,Aoki:2020bew} in the $S=-1$~\cite{Nemura:2018tay} and  $S=-2$~\cite{HALQCD:2019wsz} channels. More experimental data can be extracted from femtoscopic analyses~\cite{ALICE:2020mfd}, which allowed to constrain the $p-\Lambda$~\cite{ALICE:2018ysd}, $p-\Sigma^0$~\cite{ALICE:2019buq} and $\Lambda-\Lambda$~\cite{ALICE:2018ysd,ALICE:2019eol} interaction from HIC experimental data. 
The potentials extracted from these approaches were later used in Br\"uckner -- Hartree-Fock (BHF) calculations of hyperon potentials in ISM and pure neutron matter (NM)~\cite{halqcd2019a}, as well as the EoS~\cite{Vidana:2024ngv}. These results show that, for instance, the $\Lambda$ potential at saturation in ISM is $U_\Lambda^{\rm ISM} = -28.15 \pm 2.02 \mev$, and in NM it increases to $U_\Lambda^{\rm NM} = -25.42 \pm 1.78 \mev$. However, traditional DD-type models cannot reproduce this behavior. The single-particle potential of a baryon in the medium in RMF models is composed of the attractive scalar potential, proportional to the scalar field, and repulsive vector potential. As the composition asymmetry is increased, the magnitude of the scalar field in traditional models increases slightly~\cite{Maslov:2015wba}. However, the vector potential for $\Lambda$ does not depend on the asymmetry, because $\Lambda$ is an isoscalar and is not coupled to the isovector mesons $\rho$ or $\delta$ ($a_0(980)$). Therefore, the total $U_\Lambda^{\rm NM}$ 
decreases instead of increasing, which contradicts the microscopic results, and one of the aims of the present work is to provide a possible solution for this. Apart from modifying the EoS, the appearance of hyperons opens new channels for neutrino production via direct Urca-type processes, e.g. $\Lambda \to p + e^- + \bar \nu_e$ \cite{Maxwell:1986pj}, which lead to a rapid cooling of NSs with central density exceeding some critical value, see e.g.~\cite{Grigorian:2018bvg,Raduta:2019rsk}. For the process on $\Lambda$s mentioned above, this critical density is only slightly larger than the density of $\Lambda$ appearance in NS matter. Therefore, an accurate determination of hyperon critical densities is essential for a consistent description of the NS cooling.

In this paper, we focus on describing warm and dense isospin-asymmetric matter within the RMF framework with density-dependent (DD) couplings. Out of all possible exotic degrees of freedom, we focus on incorporating hyperons into the model. In order to describe the BHF results for hyperon potentials in NM, we extend this class of models to include not only a density dependence of the couplings, but also a dependence on the isospin asymmetry of the system. This requires breaking the SU(3) flavor symmetry in the isovector sector, while the SU(3) free parameter $z$ is used as a variable for the Bayesian analysis.
We show that the new class of models with baryon density- and isospin-density-dependent couplings and hyperons (DIDY) allows for describing the hyperon potentials in NM following from BHF calculations using a LQCD-based input~\cite{halqcd2019a}. We obtain the resulting parameterization using a Bayesian inference framework. The resulting model satisfies the constraints for saturation properties~\cite{Horowitz:2020evx,Maslov:2015wba,Lattimer:2023rpe}, NM pressure from $\chi$EFT calculations~\cite{Drischler:2021kxf}, ISM pressure from heavy-ion collisions~\cite{Oliinychenko:2022uvy}, HALQCD+BHF hyperon potentials~\cite{halqcd2019a}, and neutron star mass-radius observations from NICER~\cite{Riley:2021pdl,Miller:2021qha} and GW170817~\cite{LIGOScientific:2018cki}. The resulting general-purpose EoS, which will be made available on CompOSE~\cite{Typel:2013rza} upon publication, is supplemented with the low-density, nuclei-rich part following the well-known NSE calculation equivalent to~\cite{Hempel:2009mc}.

The paper is organized as follows. In Section~\ref{sect::model} we formulate the RMF model with density-dependent couplings and introduce the novel isospin-density dependence into them. In Section~\ref{sect::clusters} we describe the NSE model used for incorporating nuclear clusters at low densities. In Section~\ref{sect::bayesian} we discuss the Bayesian analysis performed to constrain the model parameters. In Section~\ref{sect::zeroT} we present the results for the EoS at $T = 0$ and confront the resulting NS properties with the available constraints. Finally, we illustrate the resulting EoS at $T>0$ in Section~\ref{sect::finiteT} and summarize our findings in Section~\ref{sect::conclusions}.

%%%%%%%%%%%%%%%%%%%%%%%%%%%%%%%%%%%%%%%%%%%%%%%%%%%%%%%%%%%%%%%%%%%%%%%%%%%%%%%
\section{The Model}
\label{sect::model}

%%%%%%%%%%%%%%%%%%%%%%%%%%%%%%%%%%%%%%%%%%%%%%%%%%%%%%%%%%%%%%%%%%%%%%%%%%%%%%%%%%%%%%%%%%%%%%%%%

Our model is based on the BHB$\Lambda\phi$ \cite{Banik:2014qja}
\footnote{Because this is the only model constructed by the authors and accepted on an empirical basis, we will use the alternative name ``BHB14''.} and DD2Y \cite{Marques:2017zju} models of hadronic matter, which are in turn extensions of the DD2 density-dependent relativistic mean field (DD-RMF) model \cite{Typel:2009sy}.
BHB14 introduces the $\Lambda$ hyperon, and DD2Y includes the complete ground-state baryon octet.

In the RMF framework, the forces between nucleons are mediated by the exchange of mesons. 
The $\sigma$ meson is responsible for attraction, and the $\omega$ and $\phi$ mesons for short-range repulsion, while the isovector $\rho$ introduces the isospin asymmetry, allowing for different single-particle potentials for neutrons and protons.   
The Lagrangian for this RMF model has the form \cite{Marques:2017zju}:
\begin{widetext}
\begin{equation}\label{lagrangian}
\begin{aligned}
\mathcal{L} &= \sum_{i \in B} \overline{\psi}_i (i\gamma_\mu \partial^\mu - m_i + g_{\sigma i} \sigma - g_{\omega i} \gamma_\mu \omega^\mu - g_{\phi i} \gamma_\mu \phi^\mu - g_{\rho i} \gamma_\mu \boldsymbol{\tau}_i \cdot \boldsymbol{\rho}^\mu)  \psi_i \\ 
&\phantom{=} + \frac{1}{2} (\partial_\mu \sigma \partial^\mu \sigma - m_\sigma^2 \sigma^2) - \frac{1}{4} \omega_{\mu\nu} \omega^{\mu\nu} + \frac{1}{2} m_\omega^2 \omega_\mu \omega^\mu + \frac{1}{2} m_\phi^2 \phi_\mu \phi^\mu - \frac{1}{4} \boldsymbol{\rho}_{\mu\nu} \cdot \boldsymbol{\rho}^{\mu\nu} + \frac{1}{2} m_\rho^2 \boldsymbol{\rho}_\mu \cdot \boldsymbol{\rho}^\mu,
\end{aligned}
\end{equation}
\end{widetext}
where bold symbols are vectors in isospin space, $i$ is summed over all baryons, $\psi_i$ is the Dirac spinor for baryon $i$ with corresponding mass $m_i$ summarized in Table~\ref{tab:masses}, $\gamma^\mu$ are the Dirac matrices, $\boldsymbol{\tau_i}$ is the isospin matrix (normalized so $\tau_{3} = \pm 1$ for the nucleons), $g_{Mi}$ are the density-dependent couplings of the $i$-baryon to the $M$-meson field, 
and
\begin{equation}
\begin{aligned}
\omega^{\mu\nu} &= \partial^\mu \omega^\nu - \partial^\nu \omega^\mu, \\
\phi^{\mu\nu} &= \partial^\mu \phi^\nu - \partial^\nu \phi^\mu, \\
\boldsymbol{\rho}^{\mu\nu} &= \partial^\mu \boldsymbol{\rho}^\nu - \partial^\nu \boldsymbol{\rho}^\mu.
\label{eq:vector_field_tensors}
\end{aligned}
\end{equation}
\\
are the field tensors of the Lorentz vector fields.

In the RMF formalism, the meson fields are treated as classical, and the equations of motion of the Lagrangian yield the following expectation values for the constant uniform meson fields $\sigma_0, \omega^\mu = \omega_0 \delta^{\mu 0}, \phi^\mu = \phi_0 \delta^{\mu 0}$ and $\rho_i^\mu = \rho_0 \delta^{\mu 0} \delta_{i 3}$ \cite{Typel:2009sy}:
\begin{align}
    \sigma_0 &= \sum_i \frac{g_{\sigma i}}{m_\sigma^2} n^s_i, \nonumber \\
    \omega_0 &= \sum_i \frac{g_{\omega i}}{m_\omega^2} n_i, \nonumber \\
    \phi_0 &= \sum_i \frac{g_{\phi i}}{m_\phi^2} n_i, \nonumber \\
    \rho_0 &= \sum_i \frac{g_{\rho i}}{m_\rho^2} \tau_{3i} n_i,
    \label{eom}
\end{align}
where the density of baryon $i$ is $n_i = \langle\bar{\psi}_i \gamma^0 \psi_i\rangle$, and the scalar density $n^s_i = \langle\bar{\psi}_i \psi_i\rangle$. For brevity, below we use $\sigma,\omega,\rho,\phi$ for the respective mean-field values.

%%%%%%%%%%%%%%%%%%%%%%%%%%%%%%%%%%%%%%%%%%%%%%%%%%%%%%%%%%%%%%%%%%%%%%%%%%%%%%%%%%%%%%%%%%%%%%%%%%%%%%%%%%%%%%%%%%%%%%%%%%%%%%%
\subsection{Density and isospin dependence of the coupling constants}
The original density-dependent parameterization~\cite{Typel:1999yq} was motivated by the density dependence of DBHF single-particle self-energies. However, the large-density behavior of simple parameterizations is generally not flexible enough to accommodate the saturation properties and the NS-related observables simultaneously. Within phenomenological models, a viable way of reconciling those observables is to disentangle the low-density behavior from the high-density behavior of the couplings. It can be done by adjusting the scalar field self-interaction potential, as proposed in~\cite{Maslov:2015lma}, or the dependence of the couplings on the scalar field~\cite{Maslov:2015msa,Maslov:2015wba}. In the present work, we implement a similar scheme in the model through density-dependent couplings. 

Apart from the density, the coupling constants in principle should be dependent on the isospin asymmetry. Such behavior is expected from DBHF calculations~\cite{vandalen2004,vanDalen:2005sk}, Skyrme~\cite{Cochet:2003sy} forces, and also arises naturally in some of the models with $\sigma$-dependent couplings (MKVOR-based)~\cite{Maslov:2015wba}. Phenomenologically, this is a main ingredient for fulfilling the goal of the present paper -- to describe the hyperon potentials not only in ISM, but also in NM as follows from HALQCD-based BHF calculations~\cite{Inoue:2018axd}.
We introduce an isospin dependence in the $\sigma$, $\omega$, $\rho$ and $\phi$ couplings as follows:
\begin{equation}
\begin{split}
g_{Mi} (n_B, \beta) &= \left[ 1 - \beta^2 \tanh\left(\frac{x}{e}\right)\right] g^S_{Mi}(n_B) \\&\phantom=+ \beta^2 \tanh\left(\frac{x}{e}\right) g^N_{Mi}(n_B),
\end{split}
\label{g_isospin}
\end{equation}
where $n_B = \sum_i n_i$ is the baryon density and $x = n_B/n_0$.
This form interpolates between ISM ($\beta=0$) and NM ($\beta=-1$), and extrapolates to hypothetical $\Sigma^-$- or $\Sigma^+$-rich environments with $|\beta| > 1$. The $\tanh(x/e)$ multiplier also ensures that $g_{Mi}$ is isospin-independent at zero density (necessary to ensure that $\Sigma^t$ is well-behaved as $n_B \rightarrow 0$), and we set $e = 1/3$.

The specific functional form we use in this work reads
\begin{widetext}
\begin{equation}
    g_{Mi}^{S,N}(n_B) = g^{S,N(0)}_{Mi} \left\{ \exp \left[ 1 - \left(\frac{x+1}{2}\right)^{2a_M} \right] \frac{1 - \tanh\left[(x - c_M)/d_M\right]}{2} + b_M \frac{1 + \tanh\left[(x - c_M)/d_M\right]}{2} \right\}, \label{g_density}
\end{equation}
\end{widetext}
where $g^{S,N(0)}_{Mi}$ is the coupling strength at saturation density, $a_M$ is a shape parameter for the low-density regime such that $(\partial g_{Mi}^{S,N}/\partial n_B)(n_0) \simeq -g^{S,N(0)}_{Mi} a/n_0$ for $c_M \gg 1$, $b_M$ is the ratio of the high-density coupling to $g_0$, and $c_M$ and $d_M$ are respectively the center and width of the transition zone, in units of $n_0$. As we will see below, this functional form is similar to the one in~\cite{Maslov:2015wba} after the substitution of the self-consistent scalar field.

%%%%%%%%%%%%%%%%%%%%%%%%%%%%%%%%%%%%%%%%%%%%%%%%%%%%%%%%%%%%%%%%%%%%%%%%%%%%%%%%%%%%%%%%%%%%%%%%%%%%%%%%%%%%%%%%%%%%%%%%%%%%%%%
\subsection{Coupling ratios}
The inclusion of hyperons into the model requires adjusting their respective couplings with the meson fields of the model. This is typically done using symmetry arguments for the vector mesons, while the couplings to the scalar meson $\sigma$ are deduced from the empirical value of the hyperon potentials in ISM at $n=n_0$.
The couplings of hyperons to the vector mesons in BHB14 and DD2Y  were set by SU(6) spin--flavor symmetry \cite{Banik:2014qja, Marques:2017zju}. In our work, we have reduced this symmetry to SU(3) flavor symmetry, which is physically justifiable by the ability of the $\phi$ meson to couple to the strange quark-antiquark condensate, $s\overline{s}$, in the proton \cite{Weissenborn:2011ut}. For the $\omega$ and $\phi$ mesons \cite{Marques:2017zju}, we write the couplings to the baryons in terms of the following quantities: the singlet--octet mixing angle $\theta$, the number $\alpha = F/(D+F)$ that relates the symmetric ($D$) and antisymmetric ($F$) coupling modes at vertices with three octet hadrons, and $z = g_1/g_8$, the coupling ratio between the ideal singlet and octet isoscalars. Their expressions are
\begin{align}
\begin{split}
g_{\omega N}^{S,N(0)} &= g_8 \left[ 1 - \frac{z}{\sqrt3} (1-4\alpha) \tan\theta \right],  \\
g_{\omega \Lambda}^{S,N(0)} &= g_8 \left[ 1 - \frac{2z}{\sqrt3} (1-\alpha) \tan\theta \right],\\ 
g_{\omega \Sigma}^{S,N(0)} &= g_8 \left[ 1 + \frac{2z}{\sqrt3} (1-\alpha) \tan\theta \right],\\
g_{\omega \Xi}^{S,N(0)} &= g_8 \left[  1 - \frac{z}{\sqrt3} (1+2\alpha) \tan\theta \right],\\
g_{\phi N}^{S,N(0)} &= g_8 \left[  -\tan\theta - \frac{z}{\sqrt3} (1-4\alpha) \right],\\
g_{\phi \Lambda}^{S,N(0)} &= g_8 \left[  -\tan\theta - \frac{2z}{\sqrt3} (1-\alpha) \right],\\
g_{\phi \Sigma}^{S,N(0)} &= g_8 \left[  -\tan\theta + \frac{2z}{\sqrt3} (1-\alpha) \right], \\
 g_{\phi \Xi}^{S,N(0)} &= g_8 \left[  -\tan\theta - \frac{2z}{\sqrt3} (1+2\alpha) \right].
\end{split}\label{eq:alt-couplings}
\end{align}

In the case of SU(6) symmetry, the parameters are fixed to $\theta = \arctan (1/\sqrt{2}) \approx 35.3^\circ$ (where $\phi$ becomes a pure $s\overline{s}$ state), $\alpha = 1$, and $z = 1/\sqrt{6}$. The vector mixing angle is well-constrained by the meson masses as very close to the SU(6) ideal case, so we use ideal mixing \cite{ParticleDataGroup:2024cfk}. Allowed values for the remaining two parameters are $\alpha \in [0,1]$ and $z \in [0,2/\sqrt{6}]$ \cite{Weissenborn:2011ut}. All parameter values will be fixed through Bayesian inference, as discussed in Section~\ref{sect::bayesian}. To simplify the Bayesian analysis, we choose to vary $z$ alone, while keeping $\alpha = 1$.

Instead of attempting an SU(3) extension to the scalar sector, we treat $g_{\sigma i}/g_{\sigma N}$ as free parameters, which are fitted to describe the HALQCD hyperon potentials. Some models also add the $f_0(980)$ meson (variously abbreviated as $\sigma^*$, $\sigma_s$, or $\zeta$) which couples only to hyperons and has a similar Lagrangian term and attractive force as the $\sigma$'s. However, we do not include the $f_0(980)$ in our model, and neither do BHB14 or DD2Y, because the $\Lambda \Lambda$ two-body force is poorly measured but known to be much weaker than the $\Lambda N$ force \cite{Banik:2014qja, gal2016, Tolos_2020}. For the DD2Y model, adding an $f_0(980)$ lowered the maximum NS mass from $2.04\ M_\odot$ to an unacceptable $1.87\ M_\odot$ \cite{Marques:2017zju}.

The $g_{\rho i}$ couplings are also related by SU(3); in particular, all couplings are equal in SU(6) (not including the $\tau_3$ isospin factor) \cite{oertel2015}. However, we do not include the SU(3) relations in the model, and instead vary $g_{\rho i}/g_{\rho N}$ as free parameters. Firstly, the $\rho$ coupling tends to behave differently than the $\omega$ coupling in nuclear matter, so density-dependent RMF models tend to decouple the $g_\omega$ and $g_\rho$ relations \cite{Typel:1999yq}. Also, the HALQCD potentials in NM \cite{halqcd2019a} predict a $U_\Sigma$ and $U_\Xi$ isospin splitting in NM that is much smaller than in DD2Y. Allowing $g_{\rho i}/g_{\rho N}$ to be free parameters can also overshadow the unknown coupling of the $\delta$ to hyperons, which contributes to the nucleon effective mass splitting in asymmetric matter and splits hyperon potentials in the opposite way as $\rho$; we do not include $\delta$ in our model because the nucleon mass splitting effect is poorly constrained experimentally \cite{Wang:2023owh}, and some models predict a $\delta$ coupling to nucleons which is much smaller than for the $\rho$ \cite{Santos:2024aii}.

The masses of the mesons are those from DD2Y; they are irrelevant (except in finite systems and for calibrating $\omega$ and $\phi$ couplings with SU(3)) because the couplings exclusively appear in density-dependent RMFs as $g_{Mi}/m_M^2$. For the baryons, we use the Particle Data Group mass values \cite{ParticleDataGroup:2024cfk}. All vacuum masses used for the particles included in the model are shown in Table~\ref{tab:masses}. All the coupling-related constants were calibrated with a Bayesian-inference procedure (Section~\ref{sect::bayesian}), and their values are presented in Table~\ref{tab:DID}.

\begin{table}[th!]
\vspace{.2cm}
    \centering
    \begin{tabular}{c D{.}{.}{4.3} c D{.}{.}{4.8}}
        \hline\hline
        Particle & \multicolumn{1}{c}{Mass (MeV)} & Particle & \multicolumn{1}{c}{Mass (MeV)} \\
        \hline
        $\sigma$ & 550. & $p$ & 938.272 \\
        $\omega$ & 783. & $n$ & 939.565 \\
        $\phi$ & 1020. & $\Lambda$ & 1115.683 \\
        $\rho$ & 763. & $\Sigma^+$ & 1189.37 \\
        $e^-$ & 0.511 & $\Sigma^0$ & 1192.642 \\
        && $\Sigma^-$ & 1197.449 \\
        && $\Xi^0$ & 1314.86 \\
        && $\Xi^-$ & 1321.71 \\
        \hline\hline
    \end{tabular}
    \caption{List of particles included in the model and their vacuum masses.
    }
    \label{tab:masses}
\end{table}
Charge neutrality is maintained in the model by an ideal Fermi gas of charged leptons; the Coulomb-screening effects in the jellium model are irrelevant at even the lowest density scales in our model \cite{mahan2000}. If there are both electrons and muons in cold matter, they exist in a lepton-flavor equilibrium without neutrinos. However, our model uses only electrons (with their Particle Data Group mass), since in the context of astrophysical simulations, lepton flavor equilibrium is disrupted during violent events by neutrino transport and relatively long weak decay timescales \cite{Hempel:2009mc,Banik:2014qja}. Additionally, the EoS includes a photon gas that is transparent to the medium and is described by the Stefan--Boltzmann law.

\subsection{Thermodynamics}
The grand canonical thermodynamic potential in the mean-field approximation is given by \cite{Banik:2014qja}:
%%%%%%%%%%%%%%%%%%%%%%%%%%%%%%%%%%%%%%%%%%%%%%%%%%%%%%%%%%%%%%%%%%%%%%%%%%%%%%%%%%%%%%%%%%%%%%%%%%%
\begin{widetext}
\begin{align}
\begin{split}
    \Omega/V &= \frac{1}{2} ( m_\sigma^2 \sigma^2 - m_\omega^2 \omega^2 - m_\phi^2 \phi^2 - m_\rho^2 \rho^2) - n_B \Sigma^r - \sum_{i} (\tau_{3i} - \beta) n_i \Sigma^t \\&\phantom= - \sum_i \frac{d_i T}{2\pi^2} \int_0^\infty k^2 dk \left(\ln[1 + e^{-(E^*_{ki} - \nu_i)/T}]+\ln[1 + e^{-(E^*_{ki} + \nu_i)/T}]\right),\label{Omega}
\end{split}
\end{align}
\end{widetext}
%%%%%%%%%%%%%%%%%%%%%%%%%%%%%%%%%%%%%%%%%%%%%%%%%%%%%%%%%%%%%%%%%%%%%%%%%%%%%%%%%%%%%%%%%%%%%%%%%%%%%%%%%
where $E^*_{ki}=\sqrt{k^2+{m_i^{*}}^2}$ are effective energies and $\nu_i$ are effective chemical potentials. The effective baryon mass $m_i^*$ is 
%%%%%%%%%%%%%%%%%%%%%%%%%%%%%%%%%%%%%%%%%%%%%%%%%%%%%%%%%%%%%%%%
\begin{equation}\label{mistar}
m^*_i = m_i - \Sigma^s_i = m_i - g_{\sigma i} \sigma
\end{equation}
%%%%%%%%%%%%%%%%%%%%%%%%%%%%%%%%%%%%%%%%%%%%%%%%%%%%%%%%%%%%%
and the chemical potential of the $i$th baryon is defined as
%%%%%%%%%%%%%%%%%%%%%%%%%%%%%%%%%%%%%%%%%%%%%%%%%%%%%%%%%%%%%%%%
\begin{align}
\mu_i = \nu_i +  \Sigma^v_i &= \nu_i + g_{\omega i} \omega + g_{\phi i} \phi + g_{\rho i} \tau_{3i} \rho \nonumber \\&\phantom= + \Sigma^r + (\tau_{3i} - \beta) \Sigma^t,
\label{nui}
\end{align}
%%%%%%%%%%%%%%%%%%%%%%%%%%%%%%%%%%%%%%%%%%%%%%%%%%%%%%%%%%%
where $\Sigma_i^s$ and $\Sigma_i^v$ are the scalar and vector self-energies, respectively.
Since in this approach the couplings depend on the density and isospin density, the term $n_B\Sigma^r + \sum_{i} (\tau_{3i} - \beta) n_i \Sigma^t$ is introduced in the thermodynamic potential to ensure thermodynamic consistency, where $n_B=\sum_i n_i$ is the baryon density, $\Sigma^r$ is the standard rearrangement term, and $\Sigma^t$ is a rearrangement term due to the isospin dependency of the couplings introduced in this work. The explicit expressions for these terms read

%%%%%%%%%%%%%%%%%%%%%%%%%%%%%%%%%%%%%%%%%%%%%%%%%%%%%%%%%%%%%%%%%%%%%%%%%%%%%%%%%%%%%%%
\begin{align}\label{selfr}
\begin{split}
\Sigma^r =& \sum_i \left( - \frac{\partial g_{\sigma i}}{\partial n_B} \sigma n^s_i + \frac{\partial g_{\omega i}}{\partial n_B} \omega n_i + \frac{\partial g_{\phi i}}{\partial n_B} \phi n_i \right.\\
&+ \left.\frac{\partial g_{\rho i}}{\partial n_B} \tau_{3i} \rho n_i \right),
\end{split}
\end{align}
\begin{align}\label{selft}
\begin{split}
\Sigma^t &= \frac{1}{n_B} \sum_i \left( - \frac{\partial g_{\sigma i}}{\partial \beta} \sigma n^s_i + \frac{\partial g_{\omega i}}{\partial \beta} \omega n_i + \frac{\partial g_{\phi i}}{\partial \beta} \phi n_i \right.\\
&\phantom{=\frac{1}{n_B}}+ \left.\frac{\partial g_{\rho i}}{\partial \beta} \tau_{3i} \rho n_i \right),
\end{split}
\end{align}
%%%%%%%%%%%%%%%%%%%%%%%%%%%%%%%%%%%%%%%%%%%%%%%%%%%%%%%%%%%%%%%%%%%%%%%%%%%%%%%%%%%%%%
where $n_{i}^{s}$ is the scalar density of the baryon $i$. The proof of thermodynamic consistency of the model with these rearrangement terms is presented in Appendix~\ref{app:proof}.

The single-particle potential, namely the energy gained upon addition of a baryon with 3-momentum $\mathbf{k} = 0$ to the medium, is
\begin{equation}
\begin{aligned}
U_i &= \Sigma^v_i - \Sigma^s_i \\&= -g_{\sigma i} \sigma + g_{\omega i} \omega + g_{\phi i} \phi + g_{\rho i} \tau_{3i} \rho \\&\phantom= + \Sigma^r + (\tau_{3i} - \beta) \Sigma^t.
\label{eq::ui}
\end{aligned}
\end{equation}
%%%%%%%%%%%%%%%%%%%%%%%%%%%%%%%%%%%%%%%%%%%%%%%%%%%%%%%%%%%%%%%%%%%%%%%%%%%%%%%%%%%%%%
From the thermodynamic potential, Eq. (\ref{Omega}), we can obtain all other thermodynamic quantities such as pressure, $P=-\Omega/V$. The  entropy density $s$, baryon number density $n_i$, scalar density $n_i^s$, and energy density $\epsilon$ are given by
%%%%%%%%%%%%%%%%%%%%%%%%%%%%%%%%%%%%%%%%%%%%%%%%%%%%%%%%%%%%%%%%%%%%%%%%%%%%%%%%%
\begin{align}
n_i = & \frac{\partial P}{\partial \mu_i} = \frac{d_i}{2\pi^2} \int_0^\infty k^2 dk (f_i - \overline{f}_i), \label{ni}\\
n^s_i = &\frac{\partial P}{\partial m_i^*} = \frac{d_i}{2\pi^2} \int_0^\infty k^2 dk \frac{m^*_i}{E^*_{ki}} (f_i + \overline{f}_i), \label{nis}\\
\begin{split}
\epsilon = & -P+Ts+\sum_i \mu_in_i\\
=& \frac{1}{2} \sum_M m_M^2 F_M^2 \\\phantom=& + \sum_i \frac{d_i}{2\pi^2} \int_0^\infty k^2 dk E^*_{ki} (f_i + \overline{f}_i), \label{eps}
\end{split}\\
\begin{split}
s = & \frac{\partial P}{\partial T}\\
=& \sum_i \frac{d_i}{2\pi^2} \int_0^\infty k^2 dk E^*_{ki} \left[ f_i \ln f_i + \overline{f}_i \ln \overline{f}_i\right.\\
&+ \left. (1-f_i) \ln (1-f_i) + (1-\overline{f}_i) \ln (1-\overline{f}_i) \right], \label{s}
\end{split}
\end{align}
%%%%%%%%%%%%%%%%%%%%%%%%%%%%%%%%%%%%%%%%%%%%%%%%%%%%%%%%%%%%%%%%%%%%%%%
where
\begin{subequations}
\begin{align}
    f_i(k) = \frac{1}{e^{(E^*_{ki} - \nu_i)/T} + 1},\\
    \overline{f}_i(k) = \frac{1}{e^{(E^*_{ki} + \nu_i)/T} + 1},
\end{align}
\label{eq:f_fbar}
\end{subequations}
are the Fermi distributions for baryons and antibaryons, respectively, $F_M = \sigma, \omega, \rho, \phi$ are the meson fields, and $d_{i}$ is the spin degeneracy.

We can constrain the baryon and electron chemical potentials with the four strongly conserved charges of baryon number, electric charge, strangeness, and lepton number:
\begin{align}
    \mu_i &= \mu_B + Q_i \mu_Q + S_i \mu_S \\
    \mu_e &= \mu_L - \mu_Q.
    \label{eq:mus}
\end{align}
The antibaryons are treated explicitly by means of Eqs.~\eqref{eq:f_fbar}.
In bulk matter, weak decay equilibrates the strangeness on the timescale of $\sim 10^{-10}\,\mathrm{s}$, so we can fix the hyperons by setting $\mu_S = 0$ \cite{gal2016, Banik:2014qja}. In $\beta$-equilibrium at $T = 0$, we also have $\mu_L = 0$.

\section{Description of nuclear clusters}
\label{sect::clusters}

\subsection{Low-density EoS}

\begin{figure}[h]
  \includegraphics[width=\linewidth]{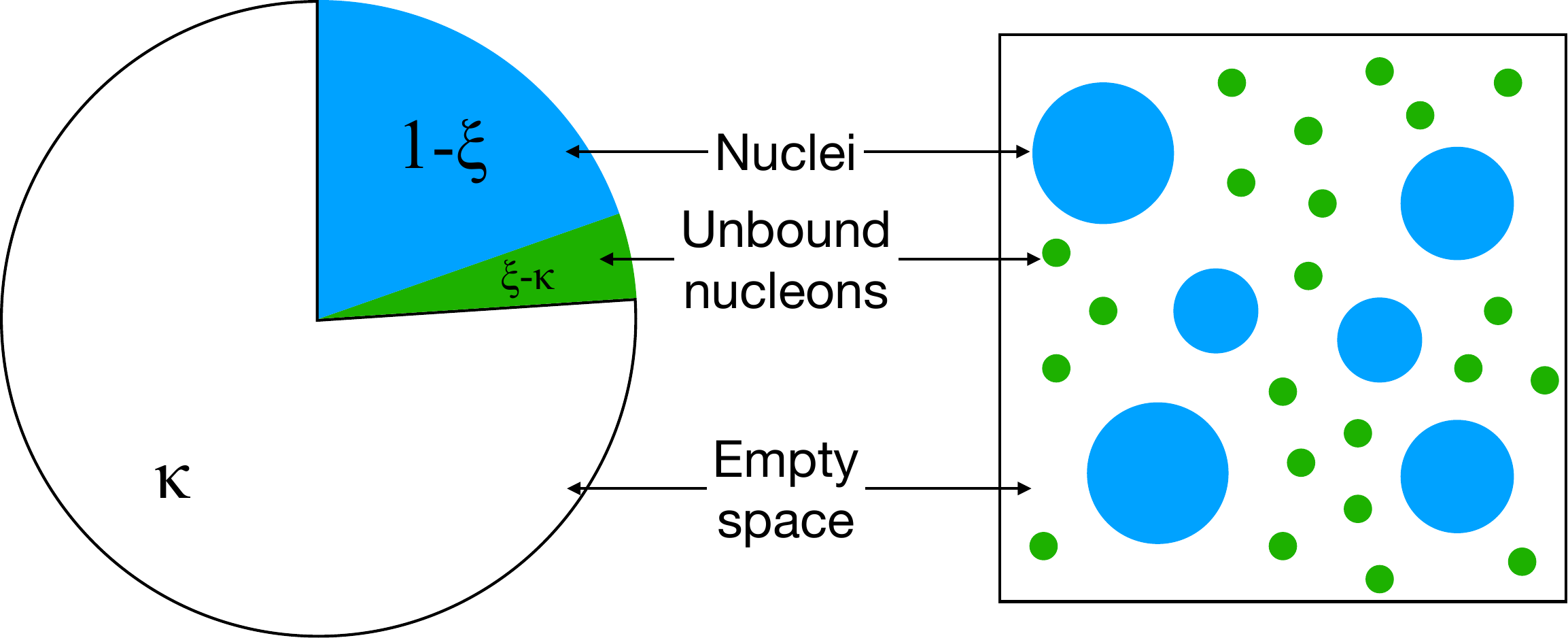}
  \caption{Cartoon representation of the HS model composition as a heterogeneous mixture of unbound nucleons (green) and bound nuclei (blue), and the corresponding fractions of space.}
  \label{volume-exclusion}
\end{figure}

In the stellar matter below the saturation density, the composition of the matter is governed by nuclear clusters. To describe their abundances and contribution to the EoS of the NS crust of and the composition of low-density nuclear matter, we used the nuclear statistical equilibrium model supplemented by the excluded-volume mechanism of cluster dissolution developed in Ref.~\cite{Hempel:2009mc}, which we will label below as HS model.
This approach assumes that below the saturation density, finite nuclei with mass number $A \geq 2$ are embedded as a non-relativistic van der Waals gas with Coulomb interactions within an RMF sea of unbound nucleons \cite{Hempel:2009mc}. The nuclei float freely in the nuclear fluid, which therefore causes a volume exclusion effect on the latter. Although the fluid is a quantum system, it can also be treated for $n_B < n_0$ as a liquid with free nucleons floating through empty space (Fig.~\ref{volume-exclusion}). In this picture, we can describe their relative volume fractions in terms of two quantities: the fraction $\kappa = 1 - n_B/n_0$ of empty space, and the fraction of space that is not occupied by nuclei,
%%%%%%%%%%%%%%%%%%%%%%%%%%%%%%%%%%%%%%%%%%%%%%%%%%%%%%%%%%%%%%%%%%
\begin{equation}
    \xi = 1 - \frac{1}{n_0} \sum_{A,Z} A n_{A,Z},
\end{equation}
%%%%%%%%%%%%%%%%%%%%%%%%%%%%%%%%%%%%%%%%%%%%%%%%%%%%%%%%%%%%%%%%%%%%%%
where $Z$ is the proton number. 
If $n_B \geq n_0$, then there are no nuclei and the HS model becomes a pure RMF.
%%%%%%%%%%%%%%%%%%%%%%%%%%%%%%%%%%%%%%%%%%%%%%%%%%%%%%%%%%%%%%%%%%%%%%%%%%
%%%%%%%%%%%%%%%%%%%%%%%%%%%%%%%%%%%%%%%%%%%%%%%%%%%%%%%%%%%%%%%%%%%%%%%
We can define $n'_i$ as the density of unbound nucleons within the space unoccupied by nuclei, so that
\begin{align}
    n_B &= \xi(n'_p + n'_n) + \sum_{A,Z} A n_{A,Z} ,\label{nb}\\
    n_p &= \xi n'_p + \sum_{A,Z} Z n_{A,Z} ,\label{np}\\
    n_n &= \xi n'_n + \sum_{A,Z} (A-Z) n_{A,Z} .\label{nn}
\end{align}
We can also calculate the baryon fractions $X_i$ for \textit{free} baryons and clusters:
\begin{align}
    X_p &= \xi n'_p / n_B ,\label{xp}\\
    X_n &= \xi n'_n / n_B ,\label{xn}\\
    X_A &= \sum_{A,Z} A n_{A,Z} / n_B. \label{xa}
\end{align}
We will also define the \textit{total} baryon fraction $Y_i$ of a specific baryon species, so that:
\begin{align}
    Y_p &= X_p + \sum_{A,Z} Zn_{A,Z} / n_B ,\label{yp}\\
    Y_n &= X_n + \sum_{A,Z} (A-Z)n_{A,Z} / n_B .\label{yn}
\end{align}
$Y_p$ is also the same as the charge fraction $Y_q$ (in the absence of hyperons). In RMF, there are no clusters, and the definitions of $X_i$ and $Y_i$ coincide.

Here, we will use the average atomic and mass number as a proxy for the overall composition of the system, such as for the various sublayers of the NS crust:
\begin{align}
\langle A \rangle = \frac{\sum_{A,Z} A n_{A,Z}}{\sum_{A,Z} n_{A,Z}} \label{ava} \, , \\
\langle Z \rangle = \frac{\sum_{A,Z} Z n_{A,Z}}{\sum_{A,Z} n_{A,Z}} \label{avz} \, .
\end{align}
Following BHB14, we include nuclei only up to $T = 50 \mev$.

% at higher temperatures, nuclei are sparse and anti-nuclei contribute to the EoS at low densities, so we also include this cutoff in our model.

\begin{figure}[h]
    \centering
    \includegraphics[width=1\linewidth]{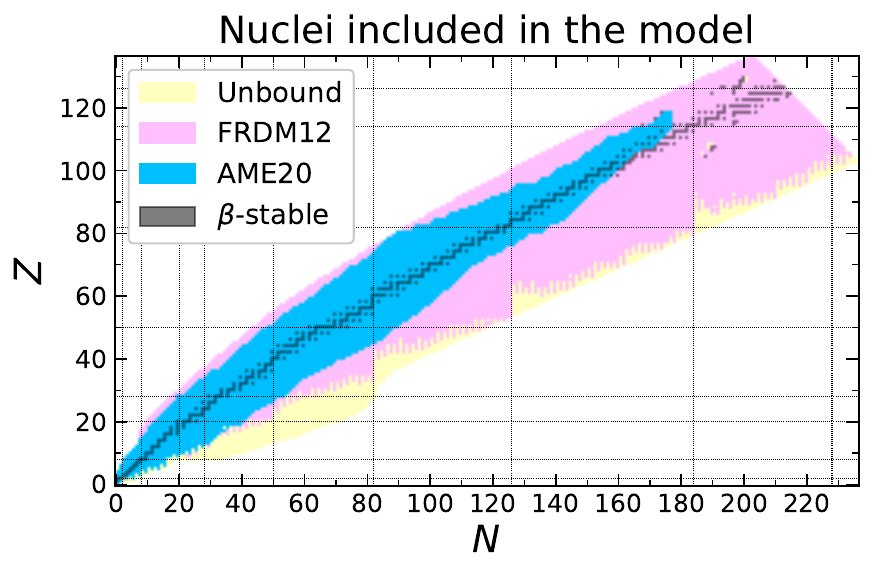}
    \caption{The table of nuclides used in our HS model; masses from AME20 are shown in blue and masses from FRDM12 are shown in pink; light-yellow nuclei are excluded by the neutron-drip cutoff. The $\beta$-stable nuclei are also shaded in dark colors.}
    \label{fig:nuclides2}
\end{figure}

We use the AME20 mass table (including systematic extrapolations, which are expected to be more accurate than theoretical models since they are extrapolated by trends on the mass surface from known nuclei with the same $Z$, $N$, $A$, or $N-Z$), together with the FRDM12 droplet model for unknown nuclei with $Z, N \geq 8$ and $A \leq 339$ \cite{Wang:2021xhn, MOLLER20161}, instead of the older AME03 and FRDM92 tables that BHB14 used \cite{Banik:2014qja}. The table ends at $A = 339$ because the model assumes that the nuclei are spherical: while the known nuclei are at most moderately deformed, this is not expected to be true for some nuclei with $A > 339$ \cite{agbemava2021}. We cut off the table at the neutron drip line, but not the proton drip line; with this exclusion, there are 8244 nuclei within our model (Fig.~\ref{fig:nuclides2}). Since the mass values in nuclear tables include electron clouds and the nuclei in HS do not have them, we apply the additional post-hoc mass correction using the semi-empirical formula from FRDM12 \cite{MOLLER20161}
%%%%%%%%%%%%%%%%%%%%%%%%%%%%%%%%%%%%%%%%%%%%%%%%%%%%%%%%%%%%%%%%%%%%%
\begin{equation}\label{d-maz}
m_{A,Z} = m_{A,Z}^0 - Z m_e - c Z^p ,
\end{equation}
%%%%%%%%%%%%%%%%%%%%%%%%%%%%%%%%%%%%%%%%%%%%%%%%%%%%%%%%%%%%
where $c = 14.33\ \mathrm{eV}$ and $p = 2.39$. However, nuclear binding energies and the position of the drip line are not subject to this correction, since the in-medium free protons also lack electron clouds.
%%%%%%%%%%%%%%%%%%%%%%%%%%%%%%%%%%%%%%%%%%%%%%%%%%%%%%%%%%%%
%%%%%%%%%%%%%%%%%%%%%%%%%%%%%%%%%%%%%%%%%%%%%%%%%%%%%%%%%%%%%%%%%%%%%%%%%%%%%%%%%%%%%%%%%%%%%%%%%%%%%%%%%%%%%%%%%%%%%%%%%%%%%%%%%
\subsection{Nuclear properties}
We treat the nuclei as spherical with nucleon density equal to $n_0$ \cite{Hempel:2009mc}. This implies a radius
\begin{equation}\label{radius}
R_{A,Z} = \left( \frac{3 A}{4\pi n_0} \right)^{1/3}.
\end{equation}
The Coulomb energy of the nucleus can be determined from the density of electrons in the surrounding medium as
\begin{equation}\label{ecaz}
\begin{aligned}
    E^c_{A,Z} &= - \frac{3}{10} \frac{Z^2 \alpha}{R_{A,Z}} (3\lambda - \lambda^3) \, , \\
    & \text{with} \; 
    \lambda = \left(\frac{n_e A}{n_0 Z}\right)^{1/3},
\end{aligned}
\end{equation}
where $\alpha \simeq 1/137$~\cite{Mohr:2024kco} is the fine-structure constant, and $n_e = Y_p n_B$ is the density of electrons in the system.

Instead of computing the large number of energy levels for each nucleus, the partition function of an individual nucleus is handled using the semi-empirical model from \cite{Fai:1982zk}:
\begin{equation}\label{gaz}
    g_{A,Z} = g_{A,Z}^0 + \frac{c_1}{A^{5/3}} \int_0^{B_{A,Z}} e^{-E/T + \sqrt{2aE}} dE,
\end{equation}
where $B_{A,Z}$ is the binding energy before applying the mass correction in Eq.~(\ref{d-maz}),
\begin{equation}
    a = \frac{A}{c_2} (1 - c_3 A^{-1/3}),
\end{equation}
$g_{A,Z}^0$ is the spin degeneracy of the ground state, $c_1 = 0.2~\mathrm{MeV}^{-1}$, $c_2 = 8~\mathrm{MeV}$, and $c_3 = 0.8$. Spin degeneracies are obtainable from NUBASE20 \cite{Kondev:2021lzi} for the empirical and systematic nuclei, and Ref.~\cite{Moller:2019jib} for the theoretical FRDM12 nuclei or when the NUBASE20 spin assignment of the ground state is unknown or uncertain (e.g.~${}^{45}\mathrm{Ar}$).

In HS, we can compute the following effective chemical potential for nuclei in terms of properties of the RMF at $n'_i$, henceforth denoted by a superscript zero:
\begin{equation}\label{nuaz}
    \nu_{A,Z} = Z\mu_p^0 + (A-Z)\mu_n^0 - m_{A,Z} - E^c_{A,Z} - P^0 V_{A,Z}.
\end{equation}
In chemical equilibrium, the number density of a nucleus is
\begin{equation}\label{naz}
    n_{A,Z} = L_{A,Z} e^{\nu_{A,Z} / T},
\end{equation}
where the coefficient is defined as
\begin{equation}\label{naznonexp}
    L_{A,Z} = \kappa g_{A,Z} \left(\frac{m_{A,Z} T}{2\pi}\right)^{3/2}.
\end{equation}
The above expression for $n_{A,Z}$ comes from setting $\mu_{A,Z} = Z \mu_p + (A-Z) \mu_n$.
%%%%%%%%%%%%%%%%%%%%%%%%%%%%%%%%%%%%%%%%%%%%%%%%%
\subsection{Thermodynamics}
In addition to the unbound nucleons, the nuclei contribute to the free energy as an ideal gas with a screened Coulomb interaction \cite{Hempel:2009mc}:
\begin{equation}
    f_{A,Z} = n_{A,Z} \left[ m_{A,Z} - T \left( 1 + \ln\frac{L_{A,Z}}{n_{A,Z}} \right) \right] + n_{A,Z} E^c_{A,Z}.
\end{equation}
The total free energy without electrons is:
\begin{equation}
    f = \sum_{A,Z} f_{A,Z} + \xi f^0(T,n'_i).
\end{equation}
The nuclei contribute to the pressure as in an ideal gas, plus the Coulomb interaction:
\begin{equation}
    P_{A,Z} = \frac{T n_{A,Z}}{\kappa} + P^c_{A,Z} ,
\end{equation}
where
\begin{equation}\label{pcaz}
    P^c_{A,Z} = -\frac{3}{10} n_{A,Z} \frac{Z^2 \alpha}{R_{A,Z}} (\lambda - \lambda^3) ,
\end{equation}
and $\lambda$ is the same as in Eq.~\eqref{ecaz}. Therefore, not including electrons,
\begin{equation}\label{p-hs}
    P = P^0(T,n'_i) + \sum_{A,Z} P_{A,Z}.
\end{equation}

The chemical potentials of the nucleons receive a correction due to the presence of the nuclei \cite{Hempel:2009mc}:
\begin{equation}
    \mu_i = \mu_i^0(T,n'_i) + \frac{1}{\kappa} \sum_{A,Z} \frac{T n_{A,Z}}{n_0}.
\end{equation}
The nuclei receive a correction as well, compared to the van der Waals gas, due to both the nucleons and the Coulomb interaction:
\begin{align}
\begin{split}
    \mu_{A,Z} =& m_{A,Z} - T \ln\frac{L_{A,Z}}{n_{A,Z}} + E^c_{A,Z} \\
    &+ V_{A,Z} \left[ P^0(T,n'_i) + \frac{T}{\kappa} \sum_{A,Z} n_{A,Z} \right].
    \label{eq:muAZ}
\end{split}
\end{align}
In chemical equilibrium, $\mu_{A,Z} = A \mu_B + Z \mu_Q$, analogous to the  baryons. The entropy of nuclei in the system is:
\begin{equation}
    s_{A,Z} = n_{A,Z} \left( \frac{5}{2} + \frac{T}{g_{A,Z}} \frac{\partial g_{A,Z}}{\partial T} + \ln\frac{L_{A,Z}}{n_{A,Z}} \right)
\end{equation}
and the total entropy follows as
\begin{equation}
    s = \xi s^0(T,n'_i) + \sum_{A,Z} s_{A,Z}.
\end{equation}
Finally, the energy density of nuclei is:
\begin{equation}
    \epsilon_{A,Z} = n_{A,Z} \left( m_{A,Z} + \frac{3}{2} T + \frac{T^2}{g_{A,Z}} \frac{\partial g_{A,Z}}{\partial T} \right) + n_{A,Z} E^c_{A,Z},
\end{equation}
so that the full energy density due to hadrons is given by
\begin{equation}
    \epsilon = \xi \epsilon^0(T,n'_i) + \sum_{A,Z} \epsilon_{A,Z}.
\end{equation}

The properties of electrons in the HS approach are identical to the ideal fermion gas, except for their chemical potential, which is corrected by the Coulomb interaction with the nuclei:
\begin{equation}
\mu_e = \mu^0_e(T, n_e) + \frac{1}{n_e} \sum_{A,Z} P^c_{A,Z},
\end{equation}
where $\mu^0_e$ is the ideal chemical potential from the RMF model. 

In principle, the HS model is fully thermodynamically consistent if the RMF model used is consistent and the table of nuclides is complete. At low temperatures, $\langle A \rangle$ and $\langle Z \rangle$ should increase without limit, and $X_A$ should approach unity, as the baryon density approaches $n_0$. However, we applied a cutoff of $A \leq 339$ on our table, so states with $\langle A \rangle > 339$ need to be approximated \cite{Hempel:2009mc}. Therefore, HS applies a Maxwell construction across the liquid-gas transition; both phases have local charge neutrality, which leads to equal $Y_q$, and are in thermal equilibrium, and the pressure $P$ and chemical potential, defined by~\cite{Hempel:2009vp}
\begin{equation}\label{mu}
\mu = (\mu_p + \mu_e) Y_q + \mu_n  (1-Y_q) = \mu_B + Y_q \mu_L \, ,
\end{equation}
are equal at both endpoints of the transition. Here $\mu_L$ follows from Eq.~\eqref{eq:mus}. Pressure and $\mu$ are constant, so $c_s^2 \equiv (\partial P / \partial \epsilon)_{s/n_B} = 0$ within the transition zone. The volume fractions of the two phases change linearly with density from the less dense phase to the more dense one.

%%%%%%%%%%%%%%%%%%%%%%%%%%%%%%%%%%%%%%%%%%%%%%%%%%%%%%%%%%%%%%%%%%%%%%%%%%%%%%%%%%%%%%%%%%%%%%%%%%%%%%%%%%%%%%%%%%%%%%%%%%%%%%%%%%%%%%%%%%%%%%%%%%%%%%%%%%%
 
\section{Bayesian inference}
\label{sect::bayesian}

We calibrated the free parameters of the model using a Bayesian inference to simultaneously reproduce hyperon
potentials from lattice QCD, nuclear matter saturation properties, and constraints from neutron star
observations and heavy-ion collisions. 
The model, which we refer to as DID (density- and isospin-dependent), and DIDY for its hyperonic extension, was
fitted using the \texttt{PyMultiNest} package \cite{pymultinest} with 2000 live points to 18 observables
described below, with likelihoods modeled as independent normally distributed variables (least-squares
minimization of $z$-scores). The maximum likelihood estimate (MLE) was then found using a minimizer initialized
at the greatest-likelihood posterior samples. While DD-RMF posteriors can contain spurious correlations, the
optimal EoS remains reliable if the functional form is well-chosen \cite{Legred:2022pyp}. The results are listed
in Table~\ref{tab:DID}.

\subsection{Parameters}

We constrained all hyperon couplings to fixed ratios relative to nucleon couplings, and similarly fixed the
ratios $g^N_{MN}/g^S_{MN}$. As follows from Eq.~\eqref{g_density}, this can be implemented by adjusting only
the saturation values $g^{S,N(0)}_{MN}$. In order to qualitatively reproduce the density-dependent coupling behavior from Ref.~\cite{vandalen2004} and ensure smooth
speed of sound profiles reminiscent of those in Ref.~\cite{Oliinychenko:2022uvy}, we have set the  transition zone parameters for vector mesons to $c_{\omega,\rho} = 3.5$ and $d_{\omega,\rho} = 1.8$ (in units of $n_0$) exempt from the Bayesian analysis.
For the $\sigma$ meson, we did not impose high-density flattening because it tends to introduce the instability $c_s^2 < 0$, which is not associated with appearance of new degrees of freedom and we ruled out as spurious. This choice of stabilizing vector meson couplings but allowing the scalar meson couplings to keep decreasing, leads to phenomenological behaviors unreachable in models with $\sigma$--field dependent couplings and hadron masses, where all couplings stop decreasing simultaneously with an increase of the density~\cite{Maslov:2015wba}. To utilize this phenomenological freedom, we fixed $c_\sigma = \infty$, which makes $d_\sigma$ irrelevant.

For convenience and lack of strong constraints on the high-density behavior of vector couplings, we performed
separate fits for five discrete values $b_\omega \in \{ 0.60, 0.65, 0.70, 0.75, 0.80 \}$, selecting the best value by requiring consistency with neutron star tidal deformability constraints from GW170817 \cite{LIGOScientific:2018hze}.
We also set $b_\rho = 0.40$ {\it a priori} because low $b_\rho$ delays hyperon onset that helps to solve both the $M_{\rm max}$ hyperon puzzle and NS cooling part of the hyperon puzzle within this class of models. 

With these constraints, there are 15 free parameters fitted to the 18 observables described below:
\begin{itemize}
	\item \textit{Nucleon $\sigma$ coupling (3 parameters).} The parameters $g^{S,N(0)}_{\sigma N}$ and $a_\sigma$ for the $\sigma$ coupling to nucleons.
	\item \textit{Hyperon $\sigma$ couplings (3 parameters).} The parameters $g^{S(0)}_{\sigma Y}$ for $Y \in \{\Lambda, \Sigma, \Xi \}$. For convenience of fitting, we varied these constants directly, not as ratios to $g^{S(0)}_{\sigma N}$. The $g^{N(0)}_{\sigma Y}$'s are in the same proportions to $g^{N(0)}_{\sigma N}$ as their ISM counterparts.
	\item \textit{Aggregated $\omega$--$\phi$ coupling (2 parameters).} The coefficient that describes the combined effects of the $\omega$ and $\phi$ in nucleonic matter \cite{Weissenborn:2011ut},
	\begin{equation}
		\tilde{g}^{S,N(0)}_{\omega N} := g^{S,N(0)}_{\omega N} \sqrt{1 + \left( \frac{g^{S,N(0)}_{\phi N} / m_\phi}{g^{S,N(0)}_{\omega N} / m_\omega} \right)^2}.
	\end{equation}
	\item \textit{Vector meson parameters (2 parameters).} The $\mathrm{SU}(3)_f$ parameter $z$ for vectors, with $\alpha = 1$, and $a_\omega$ (with $b_\omega$ surveyed over five discrete values as described above).
	\item \textit{Nucleon $\rho$ coupling (3 parameters).} The parameters $g^{S,N(0)}_{\rho N}$ and $a_\rho$ for the $\rho$ coupling to nucleons (with $b_\rho = 0.40$ fixed {\it a priori}).
	\item \textit{Hyperon $\rho$ couplings (2 parameters).} The parameters $g^{S(0)}_{\rho Y}$ for $Y \in \{\Sigma, \Xi \}$ ($g_{\rho \Lambda}$ is always zero, since the $\Lambda$ has zero isospin).
\end{itemize}

\subsection{Exclusion heuristics}

For any choice of parameters, the saturation density $n_0$ was calibrated so that $P(n_0) = 0$ in ISM without
leptons at $T = 0$. Some EoS candidates were discarded because this could not be achieved. Specifically, we
ruled out EoSs if $n_0 < 0.01 \fm^{-3}$ or $n_0 > 0.30 \fm^{-3}$, under presumption that these values are
extremely implausible or indicate absence of saturation. We used the experimental measurement $n_0 =
(0.150\pm0.010)\ \mathrm{fm^{-3}}$ from Ref.~\cite{Horowitz:2020evx}, estimated from radius measurements of
$^{208}\mathrm{Pb}$ with parity-violating electron scattering.

Since the DD couplings in DD-RMFs are not Lorentz-invariant
\cite{Typel:1999yq}, we needed to introduce the second exclusion constraint $0 < c_s^2 < 1$ at selected keypoints. We chose to evaluate this for
ISM and NM at $c_\omega n_0$, where there is a peak in the speed of sound due to the large positive second density derivative of $g_{\omega N}$ in the transition zone, which leads to an increase of $\partial \Sigma^r/\partial n_B$.

\subsection{Evidence}

The 18 observables used to constrain the model parameters are:

\begin{itemize}
\item \textit{Hyperon single-particle potentials (9 observables).} The $U_Y$ at 3-momentum $|\mathbf{k}|=0$ in ISM and NM, calculated in the BHF approach with baryon--baryon interactions obtained by HALQCD collaboration~\cite{halqcd2019a} at near-physical quark masses ($m_\pi = 146 \mev$, $m_K = 525 \mev$). This comprises three iso-multiplet averages in ISM and six individual hyperon species in NM, with statistical uncertainties $\sim \pm 2 \mev$.

\item \textit{ISM saturation properties (4 observables).} The binding energy per nucleon $B = -15.6\pm0.6 \mev$ \cite{Maslov:2015wba}, the incompressibility from GMR of $^{208}\mathrm{Pb}$ and $^{90}\mathrm{Zr}$ with $K = 240\pm20 \mev$ \cite{Maslov:2015wba}, the quadratic symmetry energy at saturation $S_2 = 32.0 \pm 1.1 \mev$ from Ref.~\cite{Lattimer:2023rpe} (fitted to various nuclear data), and the saturation density $n_0 = 0.150\pm0.010\ \mathrm{fm}^{-3}$ from Ref.~\cite{Horowitz:2020evx} (estimated from radius measurements of $^{208}\mathrm{Pb}$ with parity-violating electron scattering).

\item \textit{Finite nucleus crossover density (1 observable).} The pressure derivative
\begin{equation}
	\begin{aligned}
		M &= 3n_B \frac{\partial}{\partial n_B} \left( 9 n^2 \frac{\partial^2 B}{\partial n^2} + 18 \frac{P}{n} \right) \\
		&= x \left( 18 \frac{\partial B}{\partial y} + 12 K x + Q x^2 \right) ,
	\end{aligned}
\end{equation}
evaluated at the crossover density of $0.11 \fm^{-3} \approx 0.7\ n_0$, the average density in the outer extent of atomic nuclei where EoSs fitted to finite nuclei agree most strongly, with $M = 1100\pm70\ \mathrm{MeV}$ \cite{Khan:2012ps}.

\item \textit{NM pressure (2 observables).} While experimental constraints on NM are sparse, $\chi$EFT calculations provide reliable estimates. Ref.~\cite{Drischler:2021kxf} used an $\mathrm{N}^3\mathrm{LO}+3\mathrm{N}$ formulation to compute the pressure of NM (related to the slope of symmetry energy $L = 3n_0 \partial S / \partial n_B$ and various NS constraints~\cite{Boukari:2024wrg}) over $(0.05\text{--}0.34) \fm^{-3} \simeq (0.3\text{--}2.1)\ n_0$. We selected two keypoints to ensure well-behaved low-density matter: $n_B \in \{ 0.08,0.16 \} \fm^{-3} \simeq \{ 0.5, 1.0 \}\ n_0$.

\item \textit{Dense ISM pressure (2 observables).} To constrain dense matter behavior, we used the pressure series from Ref.~\cite{Oliinychenko:2022uvy} for ISM (excluding electrons). There, a Bayesian analysis of HIC data employed a flexible EoS parameterization with three inputs: incompressibility $K$ below $2n_0$ (with $n_0 = 0.16 \fm^{-3}$), plus two regions with constant $c_s^2$ over $(2\text{--}3) n_0$ and $(3\text{--}4) n_0$, with $c_s^2$ fixed at the near-conformal value of 0.3 above this threshold. We selected keypoints at $n_B \in \{2, 3.5 \} n_0 = \{ 0.32, 0.56 \} \fm^{-3}$. While this model likely overestimates the pressure at intermediate densities \cite{Sorensen:2023zkk}, its confidence band for $P_\mathrm{ISM}$ is consistent with DD2 and DDB.
\end{itemize}

DID is not fitted to finite nuclei, and we did not verify how accurately it describes their measured masses and shapes.

\begin{table}[h]
\centering
\begin{tabular}{c D{.}{.}{2.2} D{.}{.}{2.2} D{.}{.}{2.8} D{.}{.}{2.8}}
\hline
\multicolumn{1}{c}{Parameter} & \multicolumn{2}{c}{Range} & \multicolumn{1}{c}{MLE} & \multicolumn{1}{c}{68\% C.L.} \\\cline{2-3}
\hline
$g^{S(0)}_{\sigma N}$ & 6.00 & 11.00 & 8.94873669 & 8.263^{+0.713}_{-0.735} \\
$g^{N(0)}_{\sigma N}$ & 6.00 & 11.00 & 8.89241948 & 8.094^{+0.695}_{-0.706} \\
$a_\sigma$ & 0.00 & 1.00 & 0.16394393 & 0.189^{+0.041}_{-0.032} \\
$g^{S(0)}_{\sigma \Lambda}$ & 5.00 & 11.00 & 7.51077621 & 6.203^{+0.887}_{-0.741} \\
$g^{S(0)}_{\sigma \Sigma}$ & 3.00 & 9.00 & 6.26418057 & 4.770^{+0.999}_{-0.809} \\
$g^{S(0)}_{\sigma \Xi}$ & 1.00 & 7.00 & 6.53781517 & 4.616^{+1.309}_{-1.373} \\
\hline
$\tilde{g}^0_{\omega N}$ & 7.00 & 14.00 & 10.82857726 & 9.703^{+1.165}_{-1.255} \\
$\tilde{g}^1_{\omega N}$ & 7.00 & 14.00 & 11.00228164 & 9.698^{+1.087}_{-1.133} \\
$a_\omega$ & 0.00 & 1.00 & 0.15313180 & 0.172^{+0.048}_{-0.032} \\
$b_\omega$ & \multicolumn{2}{c}{\textemdash} & 0.80000000 & 0.800 \\
$z$ & 0.00 & \multicolumn{1}{c}{$2/\sqrt{6}$} & 0.07720445 & 0.194^{+0.135}_{-0.121} \\
\hline
$g^{S(0)}_{\rho N}$ & 0.00 & 6.00 & 3.23020263 & 3.563^{+0.400}_{-0.320} \\
$g^{N(0)}_{\rho N}$ & 0.00 & 6.00 & 2.59340047 & 2.663^{+0.644}_{-0.541} \\
$a_\rho$ & 0.00 & 4.00 & 0.39223762 & 0.212^{+0.149}_{-0.130} \\
$b_\rho$ & \multicolumn{2}{c}{\textemdash} & 0.40000000 & 0.400 \\
$g^{S(0)}_{\rho \Sigma}$ & 0.00 & 6.00 & 0.00545444 & 0.787^{+0.819}_{-0.553} \\
$g^{S(0)}_{\rho \Xi}$ & 0.00 & 6.00 & 1.11415631 & 2.096^{+0.820}_{-0.729} \\
\hline
\end{tabular}
\caption{List of parameters of DID and their values, as well as the bounds used in the Bayesian analysis.}
\label{tab:DID}
\end{table}

\begin{figure}[h]
    \centering
    \includegraphics[width=1\linewidth]{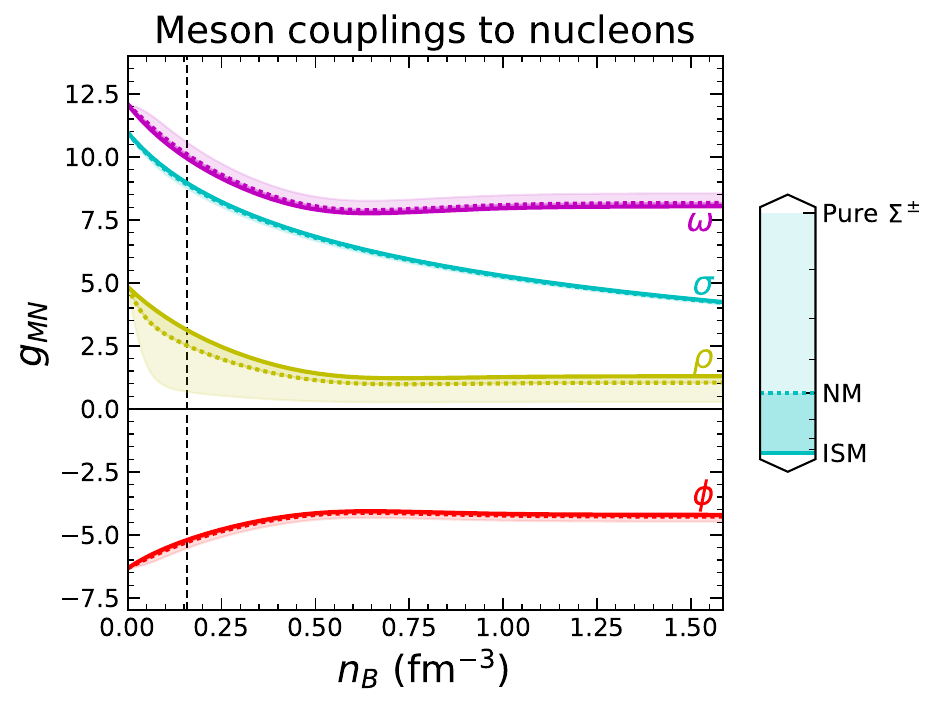}
    \caption{Plot of meson couplings to nucleons in DID as a function of baryon and isospin density. The solid lines represent ISM ($\beta=0$), the dotted lines are for NM ($\beta = -1$), and the unmarked edge of the shaded zone is the theoretical limit of $\beta = \pm2$. The vertical dashed line corresponds to the saturation density for the DID model.}
    \label{fig:couplingf}
\end{figure}

\section{Zero-temperature results}
\label{sect::zeroT}

\begin{table*}
\centering
\begin{tabular}{l c D{p}{\pm}{6.6} c D{.}{.}{4.8} D{.}{.}{4.8}}
\hline
\multicolumn{1}{c}{Quantity} & \multicolumn{1}{c}{Units} & \multicolumn{1}{c}{Empirical} & \multicolumn{1}{c}{Ref.} & \multicolumn{1}{c}{Value @ MLE} & \multicolumn{1}{c}{68\% C.L.} \\
\hline
$n_0$ & $\mathrm{fm}^{-3}$ & 0.150p0.010 & \cite{Horowitz:2020evx} & 0.15880045 & 0.153^{+0.006}_{-0.006} \\
$B$ & MeV & -15.6p0.6 & \cite{Maslov:2015wba} & -15.40 & -15.42^{+0.58}_{-0.57} \\
$K$ & MeV & 240p20 & \cite{Maslov:2015wba} & 227.06 & 212.0^{+15.0}_{-9.04} \\
$M(0.11 \fm^{-3})$ & MeV & 1100p70 & \cite{Khan:2012ps} & 1122.72 & 1132.^{+60}_{-62} \\
$S_2$ & MeV & 32.0p1.1 & \cite{Lattimer:2023rpe} & 32.44 & 32.43^{+1.06}_{-1.03} \\
\hline
$P_\mathrm{NM}(0.08 \fm^{-3})$ & $\mathrm{MeV} \fm^{-3}$ & 0.472p0.036 & \cite{Drischler:2021kxf} & 0.4569 & 0.441^{+0.032}_{-0.034} \\
$P_\mathrm{NM}(0.16 \fm^{-3})$ & $\mathrm{MeV} \fm^{-3}$ & 2.898p0.404 & \cite{Drischler:2021kxf} & 3.233 & 3.243^{+0.238}_{-0.232} \\
$P_\mathrm{ISM}(0.32 \fm^{-3})$ & $\mathrm{MeV} \fm^{-3}$ & 19.0p14.3 & \cite{Oliinychenko:2022uvy} & 12.11 & 12.36^{+0.86}_{-0.78}\\
$P_\mathrm{ISM}(0.56 \fm^{-3})$ & $\mathrm{MeV} \fm^{-3}$ & 106.8p22.0 & \cite{Oliinychenko:2022uvy} & 109.0 & 109.2^{+10.9}_{-10.7} \\
\hline
\end{tabular}
\caption{List of datapoints, other than hyperon potentials, used as evidence in the Bayesian analysis, and their MLE values and confidence intervals obtained in DID.}
\label{tab:bulks}
\end{table*}

\begin{table*}
\centering
\begin{tabular}{l D{p}{\pm}{6p4} D{p}{\pm}{6p4} D{.}{.}{3.2} D{.}{.}{3.2} D{.}{.}{5.7} D{.}{.}{5.7}}
\hline
\multicolumn{1}{c}{Quantity} & \multicolumn{2}{c}{HALQCD} & \multicolumn{2}{c}{Value @ MLE} & \multicolumn{2}{c}{68\% C.L.} \\\cline{2-3}\cline{4-5}\cline{6-7}
& \multicolumn{1}{c}{ISM} & \multicolumn{1}{c}{NM} & \multicolumn{1}{c}{ISM} & \multicolumn{1}{c}{NM} & \multicolumn{1}{c}{ISM} & \multicolumn{1}{c}{NM} \\
\hline
$\Lambda$ & -28.15p2.02 & -25.42p1.78 & -27.87 & -25.54 & -28.39^{+1.75}_{-1.69} & -25.17^{+1.52}_{-1.60} \\
\hline
$\Sigma^+$ & +14.62p1.82 & +8.24p3.68 & +14.99 & +6.85 & +15.04^{+1.68}_{-1.70} & +6.48^{+2.51}_{-2.51} \\
$\Sigma^0$ & & +15.73p1.70 & & +15.79 & & +15.67^{+1.24}_{-1.22} \\
$\Sigma^-$ & & +24.86p1.39 & & +24.74 & & +24.86^{+1.35}_{-1.33} \\
\hline
$\Xi^0$ & -3.60p2.14 & -12.19p1.46 & -3.97 & -12.13 & -3.46^{+1.68}_{-1.71} & -12.22^{+1.45}_{-1.40} \\
$\Xi^-$ & & +5.79p2.59 & & +5.85 & & +5.73^{+2.30}_{-2.35} \\
\hline
\end{tabular}
\caption{Hyperon potentials in ISM and NM at $n_B = n_0$ vs. results from \cite{halqcd2019a}. Blank cells in the ISM columns are the same across each iso-multiplet.}
\label{tab:UY}
\end{table*}

In this section, we present the results of the calibrated model at $T=0$ and compare them with existing experimental and observational constraints.
Fig.~\ref{fig:couplingf} presents the behavior of the meson couplings to nucleons as functions of baryon and isospin densities. The couplings $g_{\sigma N}$ and $g_{\omega N}$ decrease rapidly with increasing density at low values of $n_B$. Nevertheless, for $n_B\gtrsim0.5 \fm^{-3}$, $g_{\omega(\phi) N}$ saturates around $g_{\omega (\phi) N}\sim 0.7 g_{\omega (\phi) N}(0)$, while $g_{\sigma N}$ continues to decrease monotonically. Similarly, $g_{\rho N}$ declines gradually and plateaus around $0.3 g_{\rho N}(0)$ after $n_B\gtrsim0.5 \fm^{-3}$, although with a gentler slope. We also observe that the splitting due to the isospin asymmetry is quite moderate for most couplings, except for $g_{\rho N}$, where the effect is noticeably stronger.

For comparison, we also constructed DDBY, an extension of the DDB relativistic mean-field model \cite{Malik:2022zol} that uses SU(6) symmetry for vector couplings and fits $\sigma$ couplings to HALQCD-based potentials in ISM only (Table~\ref{tab:DDBY}). We used the median parameters for DDB, since Ref.~\cite{Malik:2022zol} did not report a mean or MLE for their Bayesian posterior. This EoS has saturation density $n_0 = 0.143862\ \mathrm{fm}^{-3}$.

\begin{table}
	\centering
	\begin{tabular}{c D{.}{.}{1.6}}
		\hline
		\multicolumn{1}{c}{Iso-multiplet} & \multicolumn{1}{c}{$g_{\sigma i}$}  \\
		\hline
		$N$ & 8.983000 \\
		$\Lambda$ & 5.699184 \\
		$\Sigma$ & 4.339637 \\
		$\Xi$ & 2.732038 \\
		\hline
	\end{tabular}
	\caption{The $\sigma$ couplings in DDBY. The remaining parameters of the base DDB model, except $n_0$, are listed in Ref.~\cite{Malik:2022zol}.}
	\label{tab:DDBY}
\end{table}

\subsection{Hyperon potentials}

\begin{figure*}
	\includegraphics[width=\linewidth]{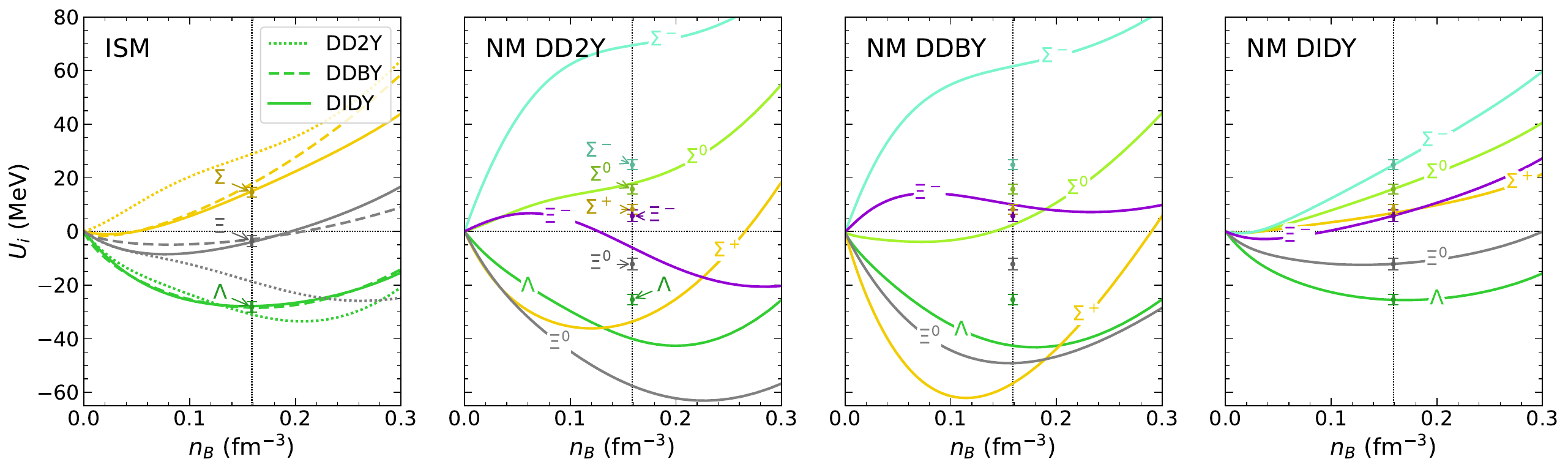}
	\caption{Left panel: Hyperon single-particle potentials as functions of baryon density $n_B$ for ISM within DDB (dashed lines) and DID (solid lines) models. Three remaining panels: Hyperon single-particle potentials in NM for DDBY and DIDY, respectively. In all panels, dots with error bars denote the BHF results~\cite{halqcd2019a} for the hyperon potentials at $|\mathbf{k}| = 0$ with their respective uncertainties.}
	\label{fig::ui}
\end{figure*}

The inclusion of hyperons into the model requires calibration of the parameters of hyperon interaction, which is most commonly done using the hyperon single-particle potential at saturation~\cite{Glendenning:1997wn}. Fig.~\ref{fig::ui} presents the hyperon single-particle potentials as functions of baryon density $n_B$ at $T=0$ for ISM (left panel), NM within the DDBY model (middle panel), and NM within the DIDY model (right panel).
In all panels, the vertical dotted line indicates $n_0$. In ISM, the DDBY (dashed lines) and DIDY (solid lines) models predict nearly identical hyperon potentials across the entire density range, as expected, because both models are fitted to describe the hyperon potentials in ISM at $n_0$. The $\Lambda$ potential remains attractive throughout, reaching approximately $-30\,\text{MeV}$ at $n_0$. The $\Sigma$ hyperon exhibits a repulsive potential, rising to $+15\,\text{MeV}$ at $n_0$, while the $\Xi$ potential shows weak attraction of about $-4\,\text{MeV}$ near saturation density.

For ISM, the only significant difference between DD2Y and our model is that the $g^{S,N(0)}_{\sigma Y}$  values are calibrated to different $U_Y$ targets \cite{Marques:2017zju}: DD2Y uses empirical estimates from Ref.~\cite{gal2016} ($-30 \mev$ for $\Lambda$, $+30 \mev$ for $\Sigma$, $-18 \mev$ for $\Xi$), while we used the LQCD-based results from Ref.~\cite{halqcd2019a}. The latter two empirical values are highly uncertain due to limited data and widely varying model estimates \cite{gal2016, halqcd2019a}.

In NM, however, the behavior of the models is dramatically different. The DDBY model (middle panel) predicts hyperon potentials that deviate significantly from the BHF results~\cite{halqcd2019a} derived from LQCD-based interactions, particularly for the $\Sigma$ and $\Xi$ hyperons. The $\Lambda$ potential for NM in DDBY model shows the trend to deepen with an increase of the asymmetry, while the LQCD-based calculation predicts a more shallow potential in NM than in ISM. This observation initially motivated our decision to develop a parameterization with isospin-dependent vector and scalar meson couplings, since $\Lambda$ hyperons are not coupled to the isovector $\rho$-meson. The $\Sigma$--hyperons, coupled to the $\rho$-meson, show isospin dependence of the potential, but it is clear that the values of the potentials from the BHF calculation are not reproduced. The isospin splitting between the $\Sigma$ species is also dramatically overestimated, which indicates that the coupling of $\Sigma$ to hyperons is too strong. For $\Xi^0$ the DDBY model overestimates the attraction at $n_0$ almost 4-fold, and $\Xi^-$ is the only species for which the DDBY model result is close to the respective BHF value. The behavior of the hyperon potentials in ISM and NM for the DD2Y model is qualitatively the same as in the DDBY model.

The DIDY model (right panel) demonstrates substantially improved agreement with the BHF constraints across all hyperon species. The $\Lambda$ potential describes well the LQCD-based data points in both ISM and NM within their uncertainties. The increased repulsion in the $\Lambda$ interaction with nucleons is attributed to the slight increase of $g_{\omega N}^{(0)}$ with an increase of the asymmetry, and the decrease of $g_{\sigma N}^{(0)}$. Since the single-particle potential in RMF models results from a subtle cancellation between large scalar and vector potentials, this change is enough to produce the increase of $U_\Lambda(n_0)$ as the asymmetry increases from ISM to NM.
The $\Sigma$ hyperon potentials show the correct splitting between isospin states. The $\Xi$ potentials also show improved agreement with BHF results, reproducing both the $\Xi^0$ and $\Xi^-$ potential values. This marked improvement in NM demonstrates that the introduction of isospin-density dependence in the coupling constants allows to consistently describe the microscopic results at different isospin asymmetries at the mean-field level.

\subsection{Particle fractions}

\begin{figure}
	\includegraphics[width=\linewidth]{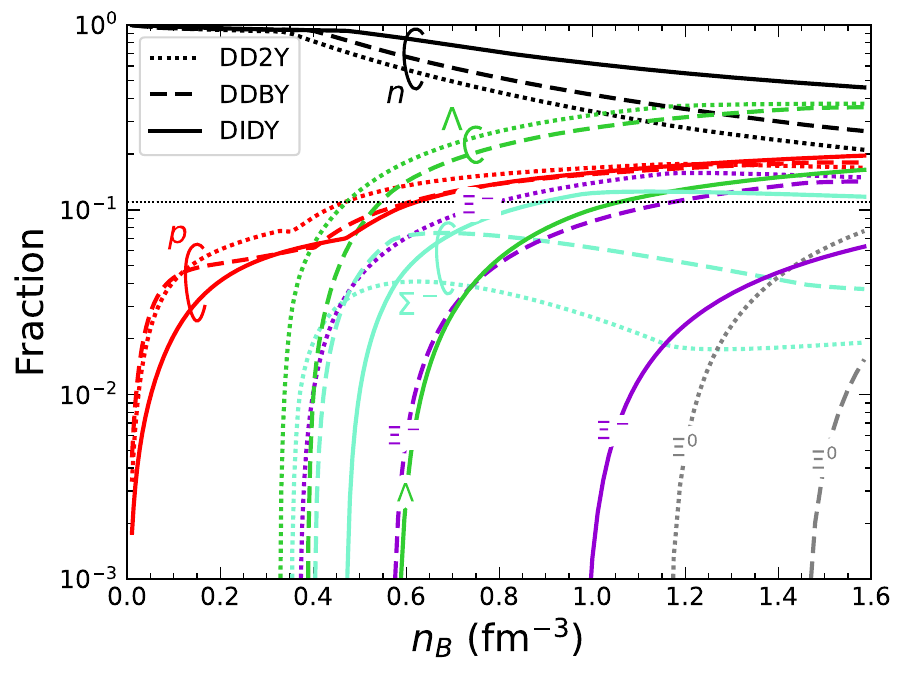}
	\caption{Baryon fractions as functions of the baryon density $n_B$ in the $\beta$-equilibrium matter with hyperons for DIDY (solid lines), DDBY (dashed lines), and DD2Y (dotted lines) models.}
	\label{fig::xi}
\end{figure}

The improved description of the hyperon potentials in the asymmetric matter directly affects the composition of dense NS matter in $\beta$-equilibrium. 
Figure~\ref{fig::xi} shows the particle fractions as functions of baryon density $n_B$ in $\beta$-equilibrium matter for two different models: DIDY and DDBY. The composition includes neutrons, protons, and hyperons $\Lambda$, $\Sigma^-$, $\Xi^-$, and $\Xi^0$ in the case of DDBY model. At low densities, matter consists primarily of neutrons with a modest proton fraction to ensure charge neutrality and $\beta$-equilibrium. 
We see that the DID model has a different proton fraction $X_p \simeq 0.03$ at $n=n_0$, while the DDB model gives $X_p \simeq 0.05$. This seeming contradiction for two models with similar saturation properties is resolved by considering the expansion of the binding energy per nucleon of nuclear matter beyond the quadratic approximation in terms of the asymmetry $\beta$. Generally, this dependence can be presented as a power series \cite{Lattimer:2023rpe}:
\begin{equation}
    B(\beta) = B(0) + S_2 \beta^2 + S_4 \beta^4 + O(\beta^6)
\end{equation}
In practice, the symmetry energy $S_2$, which we used for the Bayesian analysis, is much larger than subsequent coefficients. We can alternatively define the symmetry energy as $S \equiv B(-1) - B(0) \simeq S_2 + S_4$ (because these are inequivalent, we use a different symbol). In the parabolic approximation, where $B$ is exactly quadratic in $\beta$ and $m_e \ll \mu_B$ \cite{Harris:2025ncu},
\begin{equation}
    \mu_L = \frac{1}{n_B} \left(\frac{\partial \epsilon}{\partial X_p}\right)_{n_B} = (3\pi^2 n_B X_p)^{1/3} + 4 S_2 (2X_p - 1).
\end{equation}
Adding the next term in the expansion $B(\beta)$, of order $\beta^4$, corrects this by
\begin{equation}
    \mu_L^{(4)} = 2 \frac{\partial}{\partial\beta} ( S_4 \beta^4 ) = 8 S_4 (2X_p - 1)^3,
\end{equation}
and likewise for higher-order coefficients. Thus, a positive $S - S_2$ will lower $\mu_L$ in neutron-rich matter and raise the $X_p$ of $\beta$-equilibrium relative to the parabolic approximation ($m_e$ is still small relative to this correction). The $S_2$'s of DD2, DDB, and DID are similar ($\simeq 31 \mev$), but while $S - S_2$ is positive in DD2 and DDB, it is negative in DID (Table~\ref{tab:sms2}). Therefore, DID has a lower $\beta$-equilibrated $X_p$ at saturation density ($X_p \simeq 0.03$) than DDB and DD2 ($X_p \simeq 0.05$).

\begin{table*}[]
    \centering
    \begin{tabular}{l D{.}{.}{4.6} D{.}{.}{4.6} D{.}{.}{4.6} D{p}{\pm}{5.5} c}
        \hline
        \multicolumn{1}{c}{Quantity} & \multicolumn{1}{c}{DD2} & \multicolumn{1}{c}{DDB} & \multicolumn{1}{c}{DID} & \multicolumn{1}{c}{Emp.} & \multicolumn{1}{c}{Ref.} \\
        \hline
        $n_0$ & 0.149065 & 0.143862 & 0.158800 & 0.150p0.010 & \cite{Horowitz:2020evx} \\
        \hline
        $B$ & -16.02 & -15.27 & -15.40 & -15.6p0.6 & \cite{Maslov:2015wba} \\
        $K$ & 242.68 & 210.87 & 227.06 & 230p40 & \cite{Lattimer:2023rpe} \\
        $Q$ & 168.58 & -138.05 & -608.09 \\
        $M(0.11 \fm^{-3})$ & 1148.70 & 1215.57 & 1122.72 & 1100p70 & \cite{Khan:2012ps} \\
        \hline
        $S_2$ & 31.67 & 30.26 & 32.44 & 32.0p1.1 & \cite{Lattimer:2023rpe} \\
        $S-S_2$ & 1.00 & 0.86 & -2.72 & 1.2p1.5 & \cite{Somasundaram:2020chb} \\
        $L_2$ & 55.03 & 38.10 & 59.90 & 53p13 & \cite{Lattimer:2023rpe} \\
        $L-L_2$ & 2.67 & 2.28 & 0.05 & 0p6 & \cite{Somasundaram:2020chb} \\
        $K_{\mathrm{sym}2}$ & -93.23 & -117.99 & -130.59 & -78p43 & \cite{Lattimer:2023rpe} \\
        $K_\mathrm{sym}-K_{\mathrm{sym}2}$ & -0.46 & -0.06 & 33.27 & -24p58 & \cite{Somasundaram:2020chb} \\
        \hline
        $X_p^\mathrm{eq}(n_0)$ & 0.0523 & 0.0481 & 0.0336 \\
        \hline
    \end{tabular}
    \caption{Both definitions of symmetry energy, slope $L = 3\,\partial S/\partial x$, incompressibility $K = 9\,\partial^2 B/\partial x^2$, skewness $Q = 27\,\partial^3 B/\partial x^3$, $K_\mathrm{sym} = 9\,\partial^2 S/\partial x^2$, and $X_p$ in $\beta$-equilibrium at saturation density; as well as $M$ at $0.11 \fm^{-3}$; compared to estimates from experiment or $\chi$EFT. $X_p$ is dimensionless, $n_0$ in $\mathrm{fm}^{-3}$, and all other quantities are in MeV.}
    \label{tab:sms2}
\end{table*}

At large densities, hyperons appear with model-dependent onset densities (Table~\ref{tab:onset}). The DDBY model exhibits the conventional hierarchy, with $\Lambda$ appearing first at $n_B^{\Lambda} \simeq 0.39 \fm^{-3}$, then $\Sigma^-$ at $n_c^{\Sigma^-} \simeq 0.40 \fm^{-3}$, followed by $\Xi^-$ at $n_c^{\Xi^-} \simeq 0.57 \fm^{-3}$ and $\Xi^0$ at $n_c^{\Xi^0} \simeq 1.45 \fm^{-3}$. The $\Lambda$ fraction reaches approximately 0.28 by $n_B = 1 \fm^{-3}$. The $\Sigma^-$ fraction peaks at $n_B \simeq 0.7 \fm^{-3}$ before being displaced by $\Xi^-$, which reaches about 0.14 by $n_B \simeq 1.58 \fm^{-3}$.

\begin{table}[]
    \centering
    \begin{tabular}{l D{.}{.}{1.3} D{.}{.}{1.3} D{.}{.}{1.3}}
        \hline
        \multicolumn{1}{c}{Baryon} & \multicolumn{1}{c}{DD2Y} & \multicolumn{1}{c}{DDBY} & \multicolumn{1}{c}{DIDY} \\
        \hline
        $\Lambda$ & 0.327 & 0.386 & 0.578 \\
        $\Sigma^-$ & 0.352 & 0.401 & 0.470 \\
        $\Xi^-$ & 0.369 & 0.568 & 0.978 \\
        $\Xi^0$ & 1.162 & 1.453 & \multicolumn{1}{c}{\textemdash} \\
    \end{tabular}
    \caption{Density of hyperon onset, in $\fm^{-3}$, for cold $\beta$-equilibrated matter. The $\Xi^0$ does not appear in DIDY below the maximum density of the table, and the $\Sigma^+$ and $\Sigma^0$ do not appear in any of the models.}
    \label{tab:onset}
\end{table}

In contrast, the DIDY model inverts the first two species: $\Sigma^-$ appears first at $n_c^{\Sigma^-} \simeq 0.47 \fm^{-3}$, followed by $\Lambda$ at $n_B^{\Lambda} \simeq 0.58 \fm^{-3}$, and $\Xi^-$ at the highest density $n_c^{\Xi^-} \simeq 0.98 \fm^{-3}$. The $\Lambda$ fraction rises rapidly after onset, reaching approximately 0.16 by $n_B = 1 \fm^{-3}$, while the $\Sigma^-$ fraction peaks at $n_B \simeq 1.1 \fm^{-3}$ before being displaced by $\Xi^-$, which reaches about 0.06 by $n_B \simeq 1.58 \fm^{-3}$. The later onset densities in DIDY reflect the less attractive hyperon potentials, particularly for $\Lambda$ and $\Xi^-$.

To summarize, the DID parameterization has consistently larger critical densities for the hyperon appearance. As we will see in the next section, it leads to reduced softening of the EoS due to the hyperon appearance.

\subsection{Pressure and energy density}

\begin{figure}
	\includegraphics[width=\linewidth]{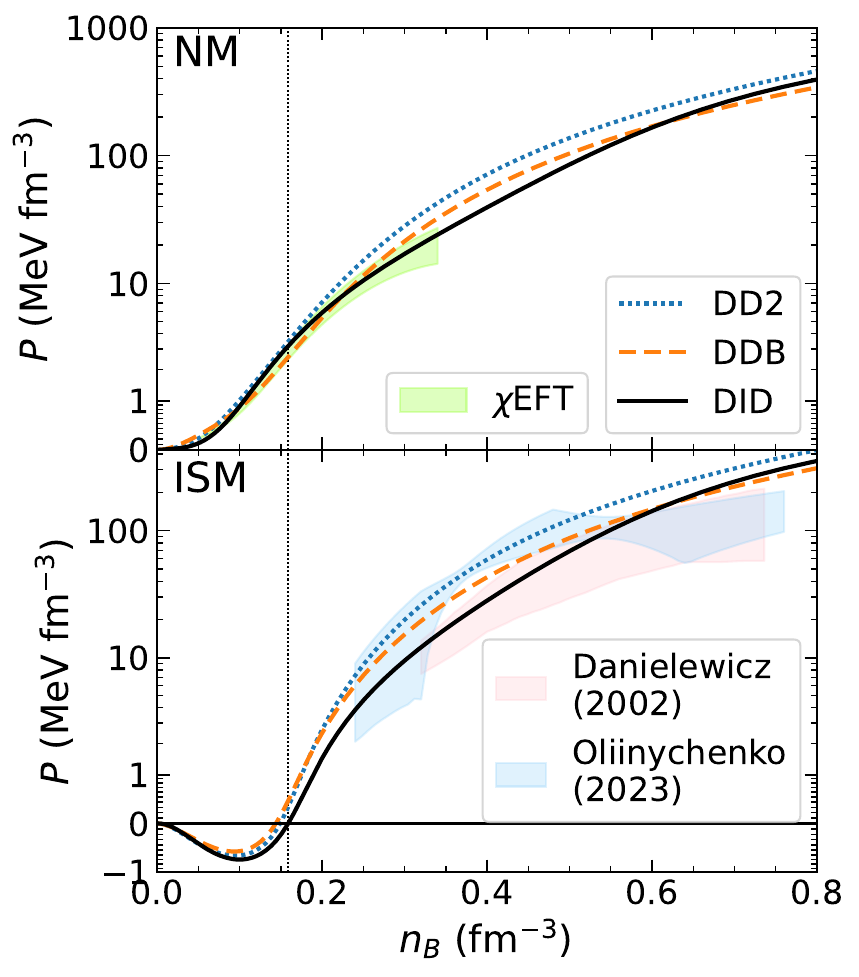}
	\caption{Pressure as a function of baryon density $n_B$ at $T=0$ for DID (solid black line), DDB (dashed orange line), and DD2 (dotted blue line) in NM (upper panel) and ISM (lower panel). The green shaded band shows the $\chi$EFT constraint~\cite{Drischler:2021kxf} for NM. The red and blue shaded bands in the lower panel show HIC flow constraints from Danielewicz et al.~\cite{Danielewicz:2002pu} and Oliinychenko et al.~\cite{Oliinychenko:2022uvy} for ISM. The vertical dotted line indicates saturation density $n_0$ for the DID model.}
	\label{fig::p_n}
\end{figure}

Fig.~\ref{fig::p_n} shows the pressure as a function of baryon density at $T=0$ for NM (upper panel) and ISM (lower panel) for the three models: the newly developed DID parameterization, the standard DD2, and DDB model. All three models have similar behavior below saturation density $n_0$, where all models provide an adequate description of the $\chi$EFT constraint~\cite{Drischler:2021kxf} for NM, with DD2 following the upper edge of the uncertainty band. For $n_B > n_0$, DD2 predicts a larger pressure than DID or DDB in both ISM and NM. For $n_B \gsim 2n_0$, DD2 significantly exceeds the $\chi$EFT constraint, predicting the stiffest EoS. Out of the three models, DID is the softest, while DDB demonstrates intermediate stiffness, remaining closer to the $\chi$EFT band than DD2. The isospin-density dependence in DID produces a moderately softer NM EoS compared to DDB in the range $n_B \simeq (1.5\text{--}3.5)\,n_0$, but afterward it overtakes it.

In ISM at low densities $n_B \lsim 0.1 \fm^{-3}$, all models exhibit negative pressure corresponding to the nuclear liquid-gas phase transition, with zero pressure at their respective saturation densities. For $n_B \gsim n_0$, the pressure has the same ordering as in NM, with DD2Y being the stiffest and DIDY the softest until it overtakes DDB near $0.60 \fm^{-3}$. Compared to HIC constraints from Refs.~\cite{Danielewicz:2002pu,Oliinychenko:2022uvy}
at $n_B \simeq (2\text{--}5)\,n_0$, corresponding to $n_B \simeq (0.3\text{--}0.8) \fm^{-3}$, DID is in good agreement with Ref.~\cite{Danielewicz:2002pu} throughout the range. Therefore, the isospin dependence of the couplings introduced to accommodate HALQCD-based hyperon potentials in neutron-rich matter does not compromise the description of ISM properties constrained by terrestrial experiments, validating the isospin-density-dependent coupling framework across the full range of isospin asymmetries.

\begin{figure}
	\includegraphics[width=\linewidth]{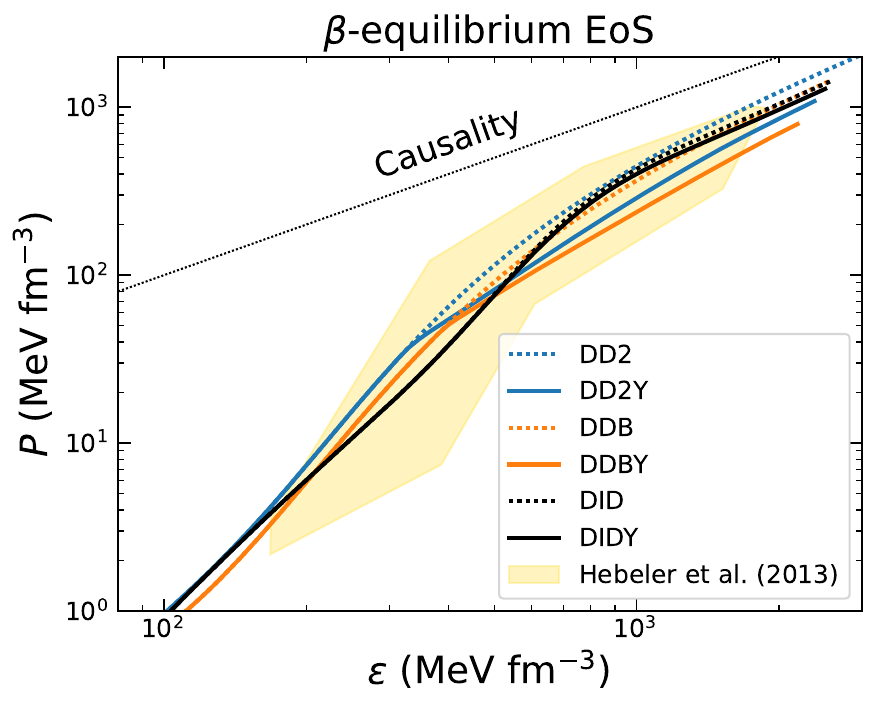}

	\caption{Pressure as a function of energy density in $\beta$-equilibrium, with and without hyperons, for DD2(Y) (blue), DDB(Y) (orange), and DID(Y) (black) models. The shaded area denotes the constraint from~\cite{Hebeler:2013nza} ($M_\mathrm{max} \geq 1.97~\,M_\odot$).}
	\label{fig::p_eps}
\end{figure}

Figure~\ref{fig::p_eps} presents the pressure as a function of energy density for $\beta$-equilibrium matter, comparing four parameterizations: the purely nucleonic models DDB and DID and their hyperonic counterparts DDBY and DIDY. At low energy densities $\epsilon \lesssim 300 \mev \fm^{-3}$, all models exhibit nearly identical behavior in the nucleon-dominated regime. All models remain consistent with the constraint band from Hebeler et al.~\cite{Hebeler:2013nza}, which incorporates causality and the requirement to support $M \simeq 1.97~M_\odot$ NSs.

At higher energy densities $\epsilon \gsim 400 \mev \fm^{-3}$, the models deviate significantly from each other, as hyperons begin to populate the matter and $g_{\omega N}$ for DID and DIDY bottom out. DDBY's early onset for hyperons causes it to soften drastically at this point compared to DDB. DID and DIDY are also moderately softer than DDB for $\epsilon \simeq (250\text{--}550) \mev \fm^{-3}$, after which the hyperon onset begins in DIDY but the EoSs are both comparable to DDB.

\begin{figure}
	\includegraphics[width=\linewidth]{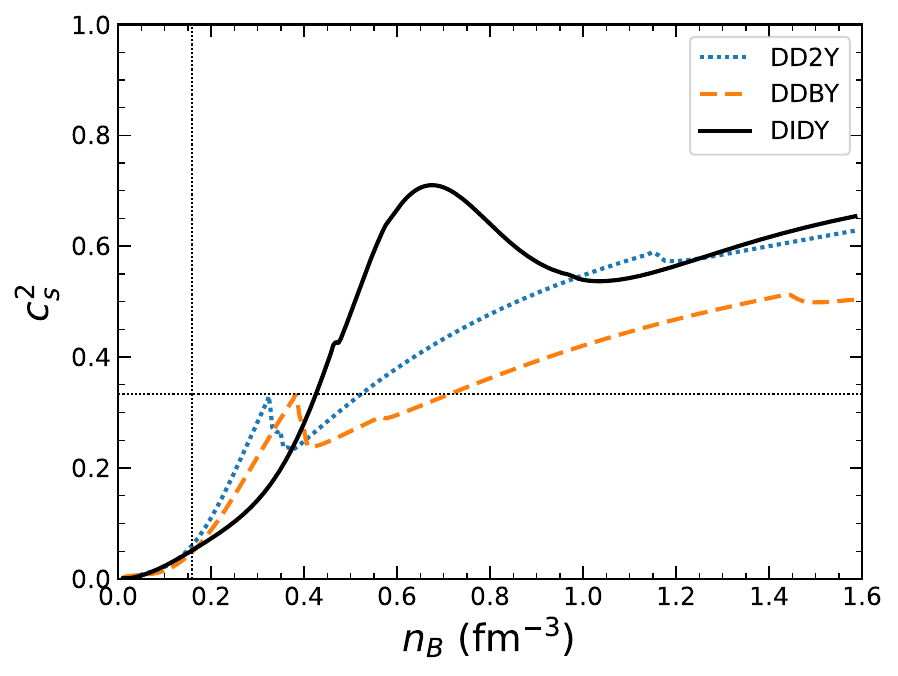}

	\caption{Speed of sound squared in $\beta$--equilibrated matter, as a function of baryon density, for DIDY (solid black), DDBY (dashed orange) and DD2Y (dotted blue) models. The horizontal dotted line denotes the conformal limit of $c_s^2=1/3$.}
	\label{fig::cs2}
\end{figure}

Figure~\ref{fig::cs2} shows the speed of sound squared $c_s^2$ as a function of baryon density $n_B$ in $\beta$-equilibrium matter for the DD2Y, DDBY and DIDY models. The DD2Y model shows a nearly monotonic increase, crossing the conformal limit (horizontal dotted line at $c_s^2 = 1/3$) at $n_B \simeq 0.55 \fm^{-3}$ and continuing to rise to $c_s^2 \simeq 0.65$ at high densities. The DDBY model displays a similar nearly monotonic increase, surpassing the conformal limit  for $n_B \simeq 0.7~\fm^{-3}$. For both these models the speed of sounds features the standard hyperon-induced sharp bends at their respective critical densities. In contrast, the DIDY model shows non-monotonic behavior with a peak of $c_s^2 \simeq 0.71$ around $n_B = 0.66 \fm^{-3}$ (approximately $4\,n_0$), followed by a decrease to a minimum $c_s^2 \simeq 0.54$ at $n_B \simeq 1.05 \fm^{-3}$, and subsequent gradual rise at higher densities that closely follows. Such non-monotonic $c_s^2$ profiles with peaks at intermediate densities are consistent with astrophysical constraints in model-independent approaches based on speed-of-sound parameterizations~\cite{Tews:2018kmu,OBoyle:2020qvf,Tan:2021ahl,Koehn:2024set,Cuceu:2024hpq}, the similar HIC-fitted constraint of \cite{Oliinychenko:2022uvy}, and in some models that include a quark phase~\cite{Rho:2024mgn,Kojo:2021ugu,Pfaff:2021kse}. The work presented in Ref.~\cite{Bedaque:2014sqa} has demonstrated that supporting 2$M_\odot$ NSs requires $c_s^2$ to exceed this bound at intermediate densities, which does happen in the DD2Y model, but at the cost of predicting large radii. RMFs do not converge to the conformal limit of $c_s^2 = 1/3$ because they lack a quark phase. For the Walecka model~\cite{Walecka1986}, which is linear with constant couplings like the asymptotic behavior of the DD-RMFs presented in this work, $c_s^2$ converges to $1$ at high density~\cite{kapusta2006}.

\subsection{Neutron star properties}

\begin{figure*}
	\includegraphics[width=.49\linewidth]{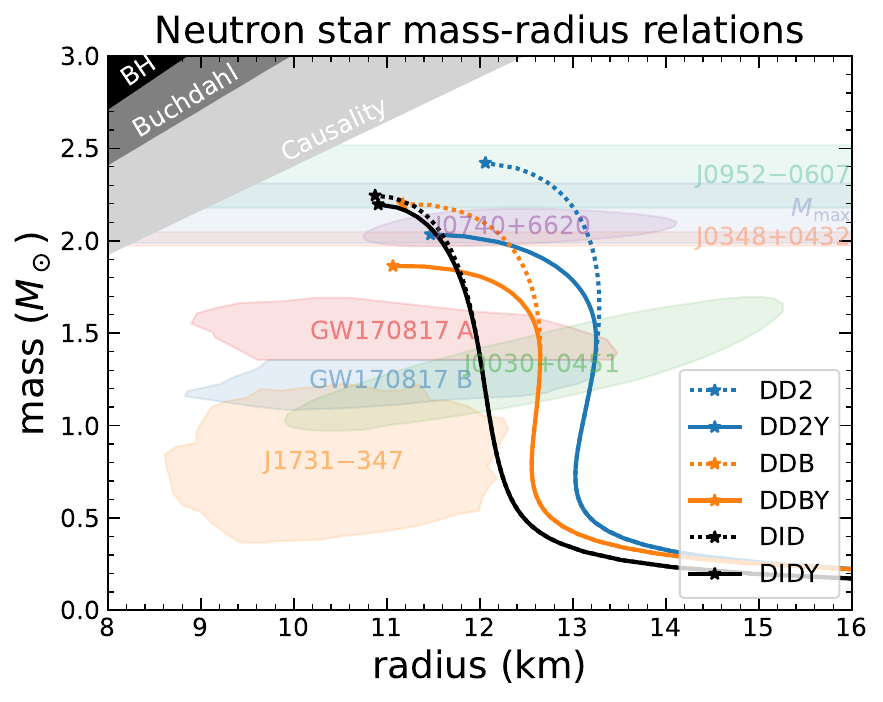}		\includegraphics[width=.49\linewidth]{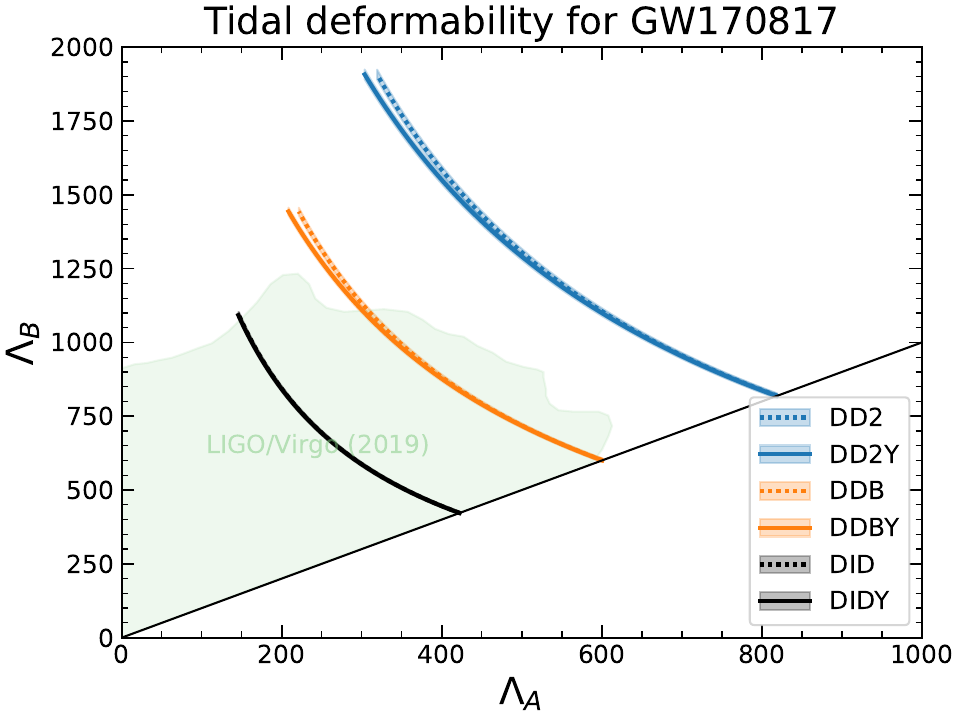}

	\caption{Left panel: Mass--radius relations for DID(Y) (black), DDB(Y) (orange), and DD2(Y) (blue) models with the HS crust with (solid lines) and without (dotted lines) hyperons. Shaded regions show observational constraints from NICER analyses of pulsars J0030+0451~\cite{Riley:2019yda} and J0740+6620~\cite{Riley:2021pdl,Miller:2021qha}, the predecessors of GW170817~\cite{LIGOScientific:2018cki}, and the low-mass compact object J1731\textminus347 \cite{Sagun:2023rzp}. Horizontal bands denote mass measurements for high-mass pulsars J0348+0432 \cite{Antoniadis:2013pzd} and J0952\textminus0607 \cite{Romani:2022jhd} and the maximum mass constraint (based on GW170817 observations) from \cite{Rezzolla:2017aly}. Gray shaded regions indicate excluded parameter space from causality and Buchdahl limits \cite{Tsokaros:2019lnx}. Right panel: $\Lambda_A$--$\Lambda_B$ relations for the predecessors of GW170817, which had a chirp mass of $1.186(1)\ M_\odot$ and a primary mass $M_A < 1.60\ M_\odot$ at 90\% confidence level \cite{LIGOScientific:2018hze}. By convention, we assume that $M_A \geq M_B$ and therefore $\Lambda_A \leq \Lambda_B$; the curves follow the permissible values of $M_A$ and $M_B$.}
	\label{fig::MR}
\end{figure*}

The left panel of Fig.~\ref{fig::MR} shows the mass--radius relations for the three models compared with modern astrophysical constraints for DD2Y, DDBY and DIDY models, calculated using QLIMR module~\cite{ReinkePelicer:2025vuh, conde_ocazionez_2025_14525356} of the MUSES calculation engine. 
We see that while the NS $M_{\rm max}$ in the DD2(Y) and DDB(Y) models are significantly affected by the appearance of hyperons, the effect of hyperons on the mass-radius curve is very small in the DID model. In particular, the decrease of the maximum NS mass is just $0.049\,M_\odot$.
Table~\ref{tab:nsprops} lists some properties of neutron stars calculated with varying EoSs and the HS crust model.

\begin{table*}
    \centering
    \begin{tabular}{l D{.}{.}{3.3} D{.}{.}{3.3} D{.}{.}{3.3} D{.}{.}{3.3} D{.}{.}{3.3} D{.}{.}{3.3}}
    \hline
    \multicolumn{1}{c}{Property} & \multicolumn{1}{c}{DD2} & \multicolumn{1}{c}{DD2Y} & \multicolumn{1}{c}{DDB} & \multicolumn{1}{c}{DDBY} & \multicolumn{1}{c}{DID} & \multicolumn{1}{c}{DIDY} \\
    \hline
    $R_{1.4}$ (km) & 13.25 & 13.24 & 12.65 & 12.65 & 11.99 & 11.99 \\
    $\Lambda_{1.4}$ & 702.37 & 700.24 & 511.30 & 510.40 & 355.13 & 355.00 \\
    $n_B^{c(1.4)}$ ($\mathrm{fm}^{-3}$) & 0.349 & 0.358 & 0.410 & 0.418 & 0.487 & 0.488 \\
    $I_{1.4}$ ($M_\odot\ \mathrm{km}^{2}$) & 86.18 & 86.10 & 79.11 & 79.07 & 71.85 & 71.85 \\
    \hline
    $M_\mathrm{max}$ ($M_\odot$) & 2.422 & 2.035 & 2.199 & 1.864 & 2.245 & 2.196 \\
    $R_\mathrm{max}$ (km) & 12.06 & 11.46 & 11.17 & 11.06 & 10.87  & 10.91 \\
    $\Lambda_\mathrm{max}$ & 6.34 & 15.13 & 6.89 & 22.30 & 4.92 & 6.13 \\
    $n_B^{c(\mathrm{max})}$ ($\mathrm{fm}^{-3}$) & 0.804 & 0.983 & 1.004 & 1.097 & 1.008 & 0.999 \\
    $I_\mathrm{max}$ ($M_\odot\ \mathrm{km}^{2}$) & 165.92 & 111.65 & 125.46 & 91.69 & 127.92 & 123.32 \\
    \hline
    \end{tabular}
    \caption{Various properties of neutron stars calculated with the HS crust model.}
    \label{tab:nsprops}
\end{table*}

The results for GW170817 tidal deformability measurements favor radii $R_{1.4} \simeq 11.5\text{--}13.5 \km$~\cite{LIGOScientific:2018cki}, while NICER observations of PSR~J0030+0451~\cite{Riley:2019yda} and PSR~J0740+6620~\cite{Riley:2021pdl,Miller:2021qha} provide direct mass-radius determinations. The massive pulsar J0740+6620 with $M = 2.08\pm0.07~M_\odot$ exceeds the maximum mass of DDBY, while DD2Y comfortably accommodates it but predicts radii near the upper observational bound for GW170817. DIDY achieves consistency with the complete set of mass-radius observations, and marginally does so for a radius measurement of the low-mass NS HESS~J1731\textminus347 \cite{Sagun:2023rzp}, while simultaneously reproducing microscopic hyperon potentials from BHF calculations. Therefore, the DIDY model offers a potentially viable solution to the hyperon puzzle through the inclusion of an isospin-dependent coupling.

In the right panel of Fig. \ref{fig::MR} we show the results for the tidal deformability parameters, $\Lambda_A$ and $\Lambda_B$, obtained from the three models and compared with the posterior distribution from Ref.~\cite{LIGOScientific:2018cki} for the binary neutron star merger GW170817, characterized by component masses $M_A \geq M_B$. The chirp mass
\begin{equation}\label{chirp}
    \mathcal{M} = \frac{(M_A M_B)^{3/5}}{(M_A + M_B)^{1/5}} = \mu^{3/5} M^{2/5}
\end{equation}
for this system is well-constrained at $\mathcal{M}=1.186(1)\, M_\odot$, since it determines the gravitational wave signal beginning at $1\frac{1}{2}$ post-Newtonian order \cite{LIGOScientific:2018cki}. For each model, the curve in the $\Lambda_A-\Lambda_B$ plane is obtained by varying $M_A$ from $2^{1/5} \mathcal{M} \approx 1.36\,M_\odot$, where the masses are equal, to the 90\% confidence upper limit of $M_A = 1.60\,M_\odot$, while fixing the chirp mass. Both the DDBY and DIDY models show good agreement with the observational constraints, whereas the DD2Y model deviates significantly from the observational constraints. However, as shown in Fig.~\ref{fig::MR}, the maximum mass predicted by the DDBY model lies below the observed lower limit of $M_{\rm{max}}\gtrsim2 M_{\odot}$. We have observed that pushing the model to lower $\Lambda_B-\Lambda_A$ values, decreases significantly the predicted maximum mass. Nevertheless, the inclusion of isospin-density-dependent couplings, together with the fine-tuning of the model parameters, allows the model to remain consistent with both the lower limits on $M_\mathrm{max}$ and the constraints on the tidal deformability parameters.

\section{Results at finite T}
\label{sect::finiteT}

\begin{figure*}
	\includegraphics[width=\linewidth]{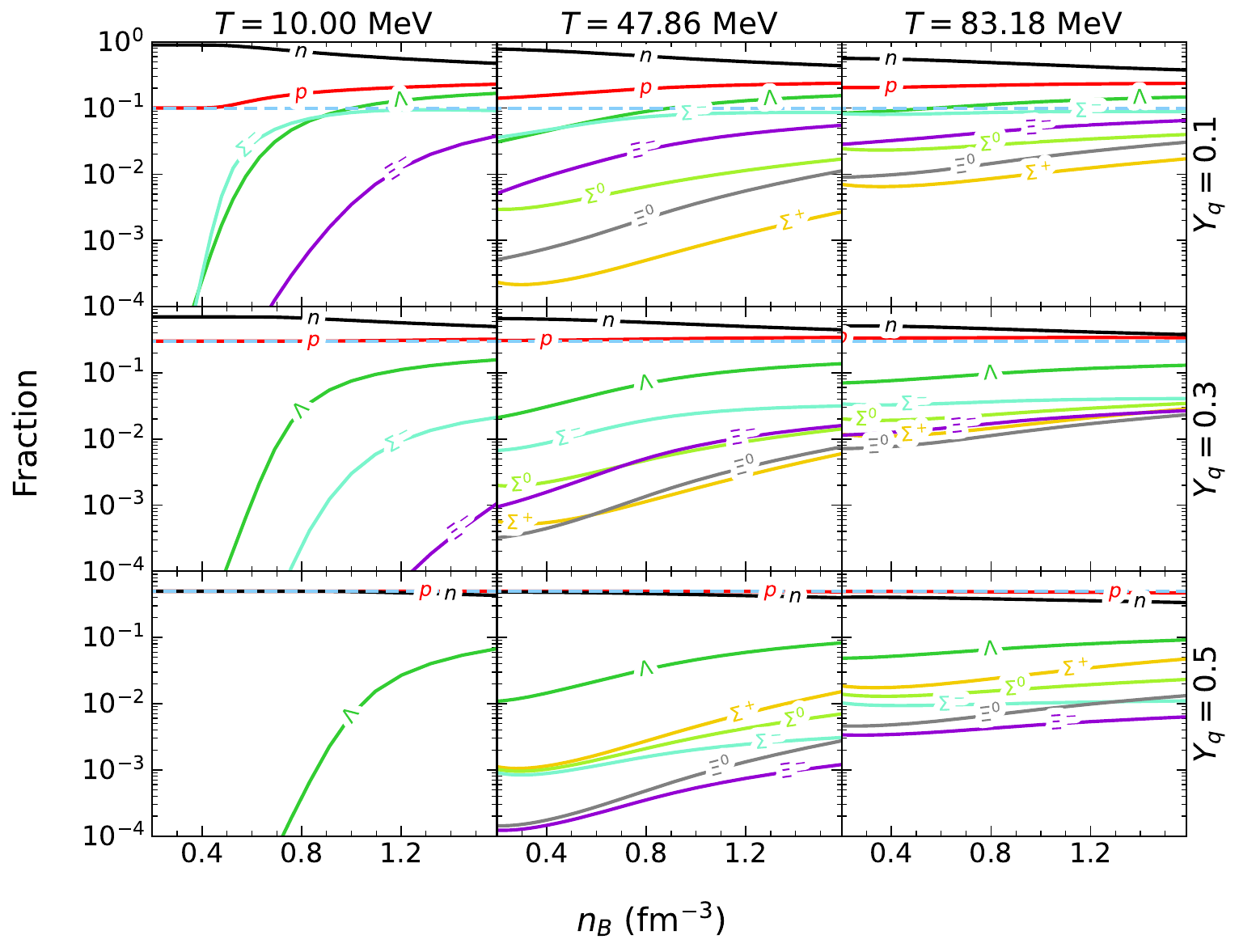}
	\caption{Plot of baryon fractions in DIDY at $T \in \left\{10^{1.00}, 10^{1.68}, 10^{1.92}\right\} \mev \approx \{ 10.00, 47.86, 83.18 \} \mev$ and $Y_q \in \{ 0.1, 0.3, 0.5 \}$. Light-blue dashed lines are electrons.}
	 \label{fig::composition_grid}
\end{figure*}

Figure.~\ref{fig::composition_grid} presents the matter composition of DIDY as a function of $n_B$ at representative temperatures entering the logarithmic grid of CompOSE $T = \{10^{1.00}, 10^{1.68}, 10^{1.92}\} \mev \simeq \{10,48,83\} \mev$ and $Y_q  = 0.1, 0.3, 0.5$. For all charge fractions, an increase in temperature has a significant impact on the matter composition, where, except for neutrons and electrons, all other species fractions increase with temperature at low densities. In the low-temperature regime ($T \lesssim 10 \mev$) the hyperon thresholds are relatively sharp. For example, $X_\Lambda$ at $T = 10 \mev$ and $Y_q = 0.3$ increases from $2.1\times10^{-4}$ at $n_B \simeq 0.52 \fm^{-3}$ to $0.075$ at $n_B = 1.00 \fm^{-3}$. Increasing the temperature to $T \simeq 48 \mev$ broadens these thresholds into smooth transitions where multiple hyperon species coexist over wider density ranges, including the $\Sigma^0$, which is strongly suppressed at low temperatures by being in chemical equilibrium with the less massive $\Lambda$.\footnote{The mass difference in vacuum is $77 \mev$, and the $\Sigma^0$ decays electromagnetically to $\Lambda$ on a timescale of $\sim 10^{-20}\,\mathrm{s}$ \cite{ParticleDataGroup:2024cfk}.} Thermal effects dominate at $T \simeq 83 \mev$, populating all hyperon species simultaneously. As the isospin asymmetry decreases from $Y_q = 0.1$ to $Y_q = 0.5$, the hyperon fractions become more isospin-degenerate, whereas for large asymmetries the negatively charged baryons have significantly larger fractions. Overall, the temperature and asymmetry dependences of the hyperon fractions are qualitatively similar to the ones observed in the literature~\cite{Tsiopelas:2024ksy}.

\section{Conclusions}
\label{sect::conclusions}

In this manuscript, we presented the first extension of the standard RMF model with density-dependent (DD) meson-baryon couplings to include an explicit isospin-asymmetry dependence. This extension was motivated by the manifest dependence of microscopic DBHF results on the isospin composition of the medium, as well as the phenomenological necessity to describe the hyperon single-particle potentials in isospin-symmetric and neutron matter in a BHF model based on HALQCD results for microscopic hyperon interactions. We parameterize the coupling dependence on the asymmetry $\beta = \sum_{i \in B} \tau_{3i} n_i$ using the leading term expansion in $\beta^2$, and perform a 17-parameter Bayesian analysis to constrain the free parameters of the model. This flexibility proved crucial for achieving simultaneous agreement with HALQCD-based hyperon potentials across media with different isospin compositions. The model is calibrated to describe saturation point observables ($n_0 = 0.150\pm0.010 \fm^{-3}$~\cite{Horowitz:2020evx}, $B = -15.6\pm0.6 \mev$, $K = 240\pm20 \mev$~\cite{Maslov:2015wba}, $S_2 = 32.0\pm1.1 \mev$~\cite{Lattimer:2023rpe}), {\it ab initio} $\chi$EFT pressure calculations for NM at sub- and near-saturation densities~\cite{Drischler:2021kxf}, phenomenological symmetric-matter pressure extracted from relativistic HIC data at suprasaturation densities~\cite{Danielewicz:2002pu}, and nine hyperon potential values ($\sim 2 \mev$ precision) from the HALQCD~+~BHF calculations~\cite{halqcd2019a}. The hyperon couplings for the $\omega$ and $\phi$ mesons follow from SU(3) flavor symmetry, for which we fix the parameters $\tan\theta = 1/\sqrt{2}$ (the $\omega$--$\phi$ mixing angle) and $\alpha = 1$ (related to antisymmetric and symmetric coupling modes at vertices with three octet hadrons), and use $z$ (the ratio between singlet and octet couplings) as a free parameter. For the isovector mesons, we treat hyperon couplings as fit parameters in order to fit the splitting between $\Sigma^{\pm, 0}, \Xi^{-,0}$ single-particle potentials in neutron matter.

The resulting parameterization, which we label as DID, reproduces all quoted observables within their uncertainty ranges. At subsaturation densities, where nuclear clusters dominate, we employ the nuclear statistical equilibrium formalism incorporating 8244 nuclei from the AME20 (experimentally measured and systematically extrapolated masses) and FRDM12 (theoretical extrapolations) mass tables with $A \leq 339$. This complements the RMF description by treating finite nuclei in thermal and chemical equilibrium with unbound nucleons, providing a transition to a pure RMF behavior at $n_B \gtrsim n_0$. In addition, it provides us with a self-consistent description of the neutron star crust in our model.

The functional shape of the coupling density dependence features a detachment of high-density from low-density behavior by means of a smooth switching construction. At low density, the parameters of the couplings follow from the Bayesian analysis, whereas at large densities the vector meson couplings are forced to become constant. This behavior is similar to the models with $\sigma$-field dependent masses and couplings~\cite{Maslov:2015wba}, where \textit{all} the couplings stop changing for densities larger than some critical value. However, in DD-type models there is no reason for all the couplings to stop decreasing, and we utilize this phenomenological freedom to not limit the decrease of the $\sigma$ coupling constants to the baryons. This phenomenological effect, providing more stiffness to the EoS, can be interpreted as a possible way to capture the manifestation of nucleon hard-core repulsion at large densities. The resulting coupling dependence on the isospin asymmetry $\beta$, necessary for describing the constraints in the DID model, is rather weak for all the isoscalar mesons. The baryon couplings to the $\rho$ meson are more affected by introducing the asymmetry dependence, which results in a more significant non-quadraticity of the symmetry energy, and accordingly decreases the proton fraction at $n_B \lsim 2\,n_0$ relative to models with near-quadratic symmetry energy at saturation.

The improvement of the description of hyperon potentials in neutron matter, introduced by the present model, has a dramatic effect on the hyperon population in the neutron star matter. Increased repulsion for the $\Lambda$ hyperon in the asymmetric matter makes it energetically less favorable to appear than the $\Sigma^-$ hyperons, and this changes the usual hierarchy of critical densities observed in other models (e.g.~\cite{Maslov:2015wba,Tsiopelas:2024ksy}) $n_c^\Lambda < n_c^{\Sigma^-} < n_c^{\Xi^-}$ to $n_c^{\Sigma^-} < n_c^\Lambda < n_c^{\Xi^-}$. This should have a significant impact on simulations of neutron star cooling, as it pushes the rather efficient direct Urca-type process on $\Lambda$ hyperons to be operative only in the largest-mass neutron stars. Moreover, in our DID model $\Lambda$ hyperons do not appear in $1.4\ M_\odot$ neutron star cores at all, which challenges the viability of efficient neutron star cooling due to the appearance of hyperons.

The resulting equation of state proves to be rather soft at intermediate densities, in accordance with the constraint from analyses of the flows in heavy-ion collisions~\cite{Danielewicz:2002pu}. Interestingly, the speed of sound $c_s^2$ within the DID model exhibits a peak for any matter composition, whereas in standard DD-type models the speed of sound keeps monotonically growing, with the exceptions of hyperon-induced sharp bends in neutron star matter. The possibility of the peak in $c_s^2$ is widely discussed in the literature as a possible signature of exotic phases in the neutron star inner core, and our DID model provides an example of a purely hadronic model which exhibits such feature. In $\beta$-equilibrium, owing to smaller hyperon abundances than in the typical RMF models, the softening effect is rather weak and leads to a decrease of the maximum predicted neutron star mass by just $0.049\,M_\odot$. This demonstrates the importance of an accurate description of hyperon potentials in neutron matter, and therefore the DID model provides a possible alternative way to resolve the hyperon puzzle. After generalization to finite temperature, the resulting general-purpose EoS spans the full thermodynamic parameter space required for modeling of neutron stars, supernovae, and binary neutron star mergers, and will be made available through the CompOSE database upon peer-review.

As an outlook, we find it extremely interesting to apply similar formalism with isospin-density-dependent couplings, deployed in this paper by the necessity to describe neutron star-related phenomenology, to a description of finite nuclei in light of the CREX/PREX controversy \cite{Lattimer:2023rpe,Miyatsu:2023lki}. The current work presents an example of large-scale Bayesian analysis, which is made possible by modern high-performance computing, and will be expanded to even larger scales upon application of machine learning-based tools like TOV emulators~\cite{Reed:2024urq}. Finally, the absence of $\Lambda$ hyperons and the proton direct Urca process in the majority of neutron stars, observed in the DID model, challenges the applicability of the ``standard cooling-like'' paradigm with hyperons to fast neutron star cooling, and supports the need for focused development of alternative scenarios.

\section*{Acknowledgments}{
This material is based upon work supported by the National Science Foundation under grants No. PHY-2208724, PHY-2116686 and PHY-2514763, and within the framework of the MUSES collaboration, under grant number No. OAC-2103680. This material is also based upon work supported by the U.S. Department of Energy, Office of Science, Office of Nuclear Physics, under Award Numbers DE-SC0022023 and DE-SC0024700 and by the National Aeronautics and Space Agency (NASA)  under Award Number 80NSSC24K0767. This work was completed in part with resources provided by the Research Computing Data Core at the University of Houston.}

\appendix
\section{Proof of thermodynamic consistency with isospin dependence of the couplings}
\label{app:proof}

	Here we show that the rearrangement terms in Eq.~(\ref{selfr}) and (\ref{selft}) make the model thermodynamically consistent.
    
Let $S$ be an index over scalar mesons and $V$ over vectors in a generic DD-RMF, and let $F_M$ be the meson fields. We will assume that $g_{Mi}$ includes the necessary $\tau_{3i}$ factor if the meson $M$ is an isovector, so that Eq.~\eqref{eom} genericizes to
\begin{equation}
\begin{aligned}
    F_S &= \sum_i \frac{g_{Si}}{M_S^2} n^s_i, \\
    F_V &= \sum_i \frac{g_{Vi}}{M_V^2} n_i.
\end{aligned}
\end{equation}

Thermodynamic consistency requires that at $T=0$, $\mu_i = \partial \epsilon / \partial n_i$ for each baryon $i$, which allows us to solve for the rearrangement terms $\Sigma^r$ and $\Sigma^t$. Then at $T = 0$, ignoring leptons:
\begin{widetext}
\begin{equation}
\begin{aligned}
\mu_i &= \sum_S m_S^2 F_S \frac{\partial F_S}{\partial n_i} + \sum_V m_V^2 F_V \frac{\partial F_V}{\partial n_i} + \sum_j \frac{\partial \epsilon_j}{\partial n_i} \\
&= \sum_{S,j} n^s_j g_{Sj} \frac{\partial F_S}{\partial n_i} + \sum_{V,j} F_V \left( g_{Vj} \frac{\partial n_j}{\partial n_i} + n_j \frac{\partial g_{Vj}}{\partial n_i} \right) + \sum_j \frac{\partial \epsilon_j}{\partial n_i} \\
&= \sum_{S,j} n^s_j g_{Sj} \frac{\partial F_S}{\partial n_i} + \sum_{V,j} F_V n_j \frac{\partial g_{Vj}}{\partial n_i} + \sum_V F_V g_{Vi} + \sum_j \frac{\partial \epsilon_j}{\partial n_i}
\end{aligned}
\end{equation}
\end{widetext}
where $\epsilon_i$ is the integral in Eq.~\ref{eps}. The derivative in $\epsilon_j$ expands as:
\begin{equation}
\begin{aligned}
\frac{\partial \epsilon_j}{\partial n_i} &= \left(\frac{\partial \epsilon_j}{\partial n_i}\right)_{m^*_j} + \frac{d_j}{2\pi^2} \int_0^{k_{Fj}} k^2 dk \frac{m^*_j}{E^*_{kj}} \frac{\partial m^*_j}{\partial n_i} \\&= \delta_{ij} \nu_j - n^s_j \sum_S \left( g_{Si} \frac{\partial F_S}{\partial n_i} + \frac{\partial g_{Sj}}{\partial n_i} F_S \right),
\end{aligned}
\end{equation}
where $k_{Fi} = (6\pi^2 n_i/d_i)^{1/3}$ is the Fermi momentum of a baryon. Substituting back in, we get:
\begin{equation}\label{nui2}
\begin{aligned}
\mu_i &= \nu_i + \sum_V F_V g_{Vi} - \sum_{S,j} n^s_j F_S \frac{\partial g_{Sj}}{\partial n_i} \\&\phantom= + \sum_{V,j} F_V n_j \frac{\partial g_{Vj}}{\partial n_i}.
\end{aligned}
\end{equation}
Since our EoS's meson couplings depend only on baryon and isospin density, each partial derivative expands as
\begin{equation}
\begin{aligned}
\frac{\partial}{\partial n_i} &= \frac{\partial}{\partial n_B} + \frac{\partial}{\partial n_i} \frac{\sum_j \tau_{3j} n_j}{n_B} \frac{\partial}{\partial \beta} \\
&= \frac{\partial}{\partial n_B} + \frac{\tau_{3i} n_B - \sum_j \tau_{3j} n_j}{n_B^2} \frac{\partial}{\partial \beta} \\
&= \frac{\partial}{\partial n_B} + \frac{\tau_{3i} - \beta}{n_B} \frac{\partial}{\partial \beta}.
\end{aligned}
\end{equation}
This means that there are two copies of the latter two terms in Eq.~\ref{nui2}: one with $\partial/\partial n_B$ and the other with $(\tau_{3i} - \beta) \partial/\partial \beta$. Comparing to Eq.~\ref{nui}, our rearrangement terms are thus those in Eqs.~(\ref{selfr}) and (\ref{selft}).

\sloppy
\bibliography{biblio.bib}

\end{document}